\documentclass[final]{elsarticle}

\usepackage[left= 3cm, right=3cm]{geometry}
\usepackage{natbib}

\usepackage{pstricks, pst-plot}	
\usepackage{graphicx} 
\usepackage{wrapfig}  
\usepackage[figuresright]{rotating}

\usepackage{amsmath}
\usepackage{amssymb}
\usepackage{amsthm}
\usepackage{hyperref}

\usepackage[english]{babel}
\usepackage{blindtext}
\usepackage{caption}
\usepackage{subcaption}
\usepackage{float}

\newtheorem{assumption}{Assumption}
\newtheorem{problem}{Problem}
\newtheorem{theorem}{Theorem}
\newtheorem{definition}{Definition}
\newtheorem{remark}{Remark}
\usepackage{algorithm}
\usepackage{algpseudocode}
\usepackage{mathtools}
\usepackage{multirow}
\usepackage{diagbox}
\begin{document}
\begin{frontmatter}%

\title{Tube-based Distributionally Robust Model Predictive Control for Nonlinear Process Systems via Linearization}
 \author[a]{Zhengang Zhong}
 \author[a]{Ehecatl Antonio del Rio-Chanona\corref{cor2}}
 \cortext[cor2]{Corresponding author}
  \author[a]{Panagiotis Petsagkourakis\corref{cor1}}
 \cortext[cor1]{Corresponding author}
  \ead{p.petsagkourakis@imperial.ac.uk}
 \address[a]{Centre for Process Systems Engineering (CPSE), Department of Chemical Engineering, Imperial College London, UK}
%
%
%
%
\begin{abstract}
Model predictive control (MPC) is an effective approach to control multivariable dynamic systems with constraints. 
Most real dynamic models are however affected by plant-model mismatch and process uncertainties, which can lead to closed-loop performance deterioration and constraint violations. Methods such as stochastic MPC (SMPC) have been proposed to alleviate these problems; however, the resulting closed-loop state trajectory might still significantly violate the prescribed constraints if the real system deviates from the assumed disturbance distributions made during the controller design.
In this work we propose a novel data-driven distributionally robust MPC scheme for nonlinear systems. Unlike SMPC, which requires the exact knowledge of the disturbance distribution, our scheme decides the control action with respect to the worst distribution from a distribution ambiguity set. This ambiguity set is defined as a Wasserstein ball centered at the empirical distribution. Due to the potential model errors that cause off-sets, the scheme is also extended by leveraging an offset-free method.
The favorable results of this control scheme are demonstrated and empirically verified with a nonlinear mass spring system and a nonlinear CSTR case study.

\end{abstract}
\end{frontmatter}
%
%
%
%
\section{Introduction}
Model predictive control (MPC) is a widely used method both in industry and academia \cite{qin2003survey, samad2020industry}, some examples include bioprocesses \cite{bradford2020stochastic, zhang2019hybrid}, reactor systems \cite{petsagkourakis2019robust, chen1995nonlinear, heirung2018stochastic}, energy storage systems \cite{kumar2019benchmarking, shirsat2021data}, and plant systems \cite{he2020modified, lu2021bayesian}. MPC's wide applications are highly motivated by its ability to explicitly deal with constraints \cite{mayne2000constrained},  such as physical or safety limitations or control design. However, due to the existence of process uncertainties in chemical systems, the processes controlled by MPC are likely to violate the prescribed constraints and hence perform poorly \cite{lorenzen2016constraint}.

To mitigate the detrimental effect of uncertainties, two sub-fields of MPC have emerged taking explicit account of uncertainties into the controller synthesis: stochastic MPC (SMPC) and robust MPC (RMPC). RMPC determines the optimal control actions with respect to the worst-case uncertainty
within a prespecified deterministic uncertainty set \cite{mayne2005robust}, whereas SMPC assumes or estimates the distribution of the uncertainty and selects the best control action for an expected objective function under soft constraints \cite{mayne2014model}. The usage of soft constraints, usually probabilistic constraints, can alleviate the conservativeness of RMPC by considering distributional information \cite{mesbah2016stochastic}.

However, in real-world applications, acquiring the true distribution of uncertainties is oftentimes challenging \cite{9222209,mark2020stochastic}. Furthermore, the high computational cost of SMPC \cite{mayne2015robust} and the discrepancy between the estimated distribution and true distribution \cite{heirung2018stochastic} limit the performance of SMPC for practical applications.

To address these challenges of SMPC - conservativeness, computational complexity, and estimation of distribution - we propose a data-driven distributionally robust MPC (DRMPC) for nonlinear systems using the Wasserstein metric. In the construction of this controller, instead of knowing the probabilistic distribution of disturbances exactly, only samples of the disturbance realizations are required. These samples convey the partially known distributional information, and are used to construct the Wasserstein ambiguity set. This ambiguity set contains the empirical distribution at its centre, and includes all  distributions within some distance from this empirical distribution of collected samples, the distance is in the Wasserstein sense. Additionally, the resulting distributionally robust (DR) controller exhibits behaviour between a SMPC and a RMPC by simply scaling a single parameter: the size of the Wasserstein ambiguity set. Furthermore, to alleviate the computational complexity, this optimal control problem is reformulated into a conic program with the help of convex analysis.


The idea of applying ambiguity sets into MPC originates from the distributionally robust optimization (DRO) problems. DRO has remarkably developed as an optimization framework in the past decade because of its ability to explicitly take the ambiguity of the underlying probability distribution into account \cite{rahimian2019distributionally}. Furthermore, DRO as a generalization of both robust optimization (RO) and stochastic optimization (SO) is able to resolve their inherent limitations such as conservativeness, poor out-of-sample performance, and computational intractability \cite{sun2021robust, mohajerin2018data}. As the core of DRO design, the ambiguity set is a set of distributions over which the worst distribution is to be determined. Throughout this paper, we consider distance-based ambiguity sets: data-driven Wasserstein set \cite{mohajerin2018data}

DRO-based optimal control has gained growing attraction since the early work of \cite{van2015distributionally} specifying a non-data-driven moment-based ambiguity set with known first two moments of the disturbance distribution. Another work using the moment-based ambiguity set \cite{lu2020soft} considers the first-order moment information. 
In the context of data-driven control, \cite{coppens2020data} considers multiplicative noise complying with sub-Gaussian distributions, \cite{schuurmans2021general,schuurmans2021data} restrict their attention to Markov jump linear systems, and \cite{yang2020wasserstein} proposes a dynamic programming solution for Wasserstein DR control problems without imposing state and input constraints.

This work is extended from the Authors' previous work \cite{zhong2021data} on linear systems. Instead of applying the feedback policy in terms of disturbance, we reformulate the problem into a tube-based setting to reduce the computational complexity. Also we provide two linearization methods to deal with the nonlinearity. Two other closely relevant papers are \cite{coulson2021distributionally} and \cite{mark2021data}. They both propose DR control frameworks for linear systems with unbounded additive disturbances by imposing data-driven DR Wasserstein chance constraints on system states and reformulating optimization problems into tractable forms.

We summarize our contributions as follow. (i) We propose a general data-driven DRMPC scheme for nonlinear systems with additive disturbances such that the prescribed expected state constraints and robust input constraints can be satisfied; (ii) The proposed scheme can be reformulated into a conic program through successive linearization or feedback linearization, and the inherited offset is accounted for; (iii) The proposed algorithm progressively increases constraint satisfaction as more samples are collected; (iv) Simulation results of nonlinear systems are illustrated to verify the functionality of the proposed distributionally robust state constraints and compared with polynomial-chaos-based SMPC, all code is made available for reproducibility.

\section{Problem Statement}
\subsection{Notations}
We use $x_{k}$ for the measured state at time $k$ and $x_{i \mid k}$ for the state predicted $i$ steps ahead at time $k$. $[A]_{j}$ and $[a]_{j}$ denote the $j$-th row and entry of the matrix $A$ and vector $a$, respectively. Similarly we denote the element of i-th row and j-th in the matrix $A$ as $[A]_{ij}$. We also define the notation $[A]_{i:j}$ for the i-th row to j-th row in the matrix $A$. The set $\mathbb{N}_{>0}$ denotes the positive integers and $\mathbb{N}_{\geq 0}=\{0\} \cup \mathbb{N}_{>0}$. The notation $\mathbb{E}_{k}\{\mathcal{A}\}=\mathbb{E}\left\{\mathcal{A} \mid x_{k}\right\}$ denotes the conditional expectation of an event $\mathcal{A}$ given the realization $x_k$. $\mathcal{M}(\Xi)$ defines the space of all probability distributions supported on $\Xi$ with finite first moments. $(\cdot)^{(i)}$ denotes the i-th sample from the training set. The sequence of length $N$ of vectors $v_{0 \mid k}, \ldots, v_{N-1 \mid k}$ is denoted by $\mathbf{v}_{N \mid k}$. $I_i$ denotes a column vector, in which only the i-th entry is $1$ and the remaining entries are $0$. $\gamma_{ijl}$ denotes the element of a 3-D tensor, such that this element is the i-th, j-th, l-th element along the first, second and third axis, respectively. Similar for $t_i$ and $\xi_{ij}$ in 1-D and 2-D, respectively.

\subsection{System dynamics, Constraints and Objective}
We consider the nonlinear time-invariant stochastic dynamical system with additive disturbance
\begin{equation}
\label{eq:system}
x_{k+1}=f(x_{k}, u_{k})+ Dw_{k}, \quad k \in \mathbb{N}_{\ge 0},
\end{equation}
where $k$ is the discrete time and $f: \mathbb{R}^{n_{x}} \times \mathbb{R}^{n_{u}} \times \mathbb{R}^{n_w} \rightarrow \mathbb{R}^{n_{x}}$ denotes the known nonlinear system dynamics with the state $x_{k} \in \mathbb{R}^{n_x}$, the control $u_{k} \in \mathbb{R}^{n_u}$, and the additive disturbance $w_{k} \in \mathbb{R}^{n_w}$. Each disturbance $w_k$ of the disturbance sequence $\{ w_{k} \}_{k \in \mathbb{N}_{\ge 0}}$ is assumed to be a realization of the corresponding random variable (r.v.) $W_{k}$ from the random process $\{W_{k}\}_{k \in \mathbb{N}_{\ge 0}}$ satisfying the following assumption.
\begin{assumption}[Bounded i.i.d Random Disturbance]
\label{assump:iid}
All random variables $W_k$ for $k \in \mathbb{N}_{\ge 0}$ from the family of random variables $\{W_{k}\}_{k \in \mathbb{N}_{\ge 0}}$ are assumed to be independent and identically distributed (i.i.d) with an unknown probability distribution $\mathbb{P}_{w}$ and the polyhedral support $\mathbb{W}_{w} \triangleq \{w \mid H_w w \le h_w\}$.
\end{assumption}

For any given state measurement $x_k$ at the sample time $k$, the predicted system states are described as 
$$
\footnotesize
x_{i+1 \mid k}=f(x_{i \mid k}, u_{i \mid k})+D W_{i+k} \quad x_{0 \mid k} \stackrel{a . s .}{=} x_{k},
$$
where $x_{i \mid k}$ and $u_{i \mid k}$ are both random variables. The sequence of random variables $\{W_{i}\}_{i \in [k, k+N-1]}$ from $k$ to $k+N-1$ within the prediction horizon $N$ is denoted as $\Xi_{k}$ and the corresponding realization is denoted as $\xi_{k}$. Due to Assumption \ref{assump:iid}, $\Xi_{k}$ $\forall{k \in \mathbb{N}_{k \ge 0}}$ complies with the same product distribution $\mathbb{P}_{\xi} \triangleq \underbrace{\mathbb{P}_{w} \times \dots \times \mathbb{P}_{w}}_{N}$. Hence we can denote the disturbance realizations as $\xi$ unambiguously by dropping the subscription $k$ and thereby the polyhedral support $\{ \mathbb{W}_{\xi} \mid H_{\xi} \xi \le h_{\xi} \}$. 

For any nonlinear system, we consider distributionally robust constraints on the states and hard constraints on the inputs:
\begin{equation}
\label{eq:proto_constraint}
\begin{aligned}
\sup_{\mathbb{P}_k \in \mathcal{P}_k} \mathbb{E}_{\mathbb{P}_k}\left\{[F]_{j}x_{i \mid k}\right\} & \leq [f]_{j}, \quad k \in \mathbb{N}_{\ge 0}, j \in \mathbb{N}_1^{n_F}, i \in \mathbb{N}_1^N \\
G u_{i \mid k} & \leq g \,  \quad  i \in \mathbb{N}_0^{N-1},
\end{aligned}
\end{equation}
where $F \in \mathbb{R}^{n_F \times n_x}, \quad G \in \mathbb{R}^{n_G \times n_u}, \quad f \in \mathbb{R}^{n_F}, \quad g \in \mathbb{R}^{n_G}$. Here $\mathcal{P}_{k}$ is the ambiguity set constructed as the Wasserstein ball centralized around the empirical distribution $\hat{\mathbb{P}}_{k}:=\frac{1}{N_{k}} \sum_{l=1}^{N_{k}} \delta_{\hat{\xi}^{(l)}}$. By imposing the constraints mentioned above, we could guarantee that the nominal state constraint for the closed-loop system
\begin{equation}
\label{eq:nom_state_constraint}
Fx_{k} \le f
\end{equation}
can be satisfied with high probability. More details will be introduced in Section \ref{sec:DRO_ambiguity}.

Without loss of generality, we characterize the control target as tracking the equilibrium point, which we assume to be the origin of the coordinate system, from an initial state while satisfying the prespecified constraints. The control objective function can hence be defined as the minimization of the expected value with the reference trajectories uniformly equal to zero
\begin{equation}
\label{eq:proto_obj}
\footnotesize
\mathbb{E}_{\mathbb{P}}\left\{\sum_{i=0}^{N-1}(\left\|x_{i \mid k}\right\|_{Q}^{2}+\left\|u_{i \mid k}\right\|_{R}^{2}) + \left\|x_{N \mid k}\right\|_{Q_f}^{2}\right\}.
\end{equation} Here $Q, Q_f \in \mathbb{R}^{n_x \times n_x}$ and $\mathbb{R}^{n_u \times n_u} $ are penalty matrices for the quadratic stage costs. The corresponding optimization problem of DRMPC for nonlinear systems is defined as 

\begin{problem}
\label{prob:orig_prob}
\begin{equation}
\label{eq:prototype_prob}
\begin{array}{cl}
\displaystyle\min_{\mathbf{u}} & \mathbb{E}_{\mathbb{P}}\left\{\sum_{i=0}^{N-1}(\left\|x_{i \mid k}\right\|_{Q}^{2}+\left\|u_{i \mid k}\right\|_{R}^{2}) +\left\| x_{N \mid k}\right\|_{Q_f}^{2}\right\} \\
\text { s.t. } 
& x_{0\mid k}=x_{k}\\
& x_{i+1 \mid k}=f(x_{i \mid k}, u_{i \mid k}) + D W_{k+i}\\
& \displaystyle\sup_{\mathbb{P}_k \in \mathcal{P}_k} \mathbb{E}_{\mathbb{P}_k}\left\{[F]_{j}x_{i \mid k}\right\}  \leq [f]_{j}, \quad k \in \mathbb{N}_{\ge 0}, j \in \mathbb{N}_1^{n_F}, i \in \mathbb{N}_1^N \\
& G u_{i \mid k}  \leq g \,  \quad  i \in \mathbb{N}_0^{N-1}.\\
\end{array}
\end{equation}
\end{problem}

We propose a successive linearization method to approximately solve the proposed DRMPC \eqref{eq:prototype_prob} in Section \ref{sec:suc_lin}, such that the closed-loop system can comply with the nominal constraints \eqref{eq:nom_state_constraint}. For a special form of nonlinear systems, i.e. control-affine systems, we applied feedback linearization to solve the proposed DRMPC in Section \ref{sec:feed_lin}. 

\section{Distributionally robust optimization and Wasserstein Ambiguity Sets}
\label{sec:DRO_ambiguity}
\subsection{Distributionally robust optimization}
Distributionally robust optimization is an optimization model which utilizes the partial information about the underlying probability distribution of the random variables in a stochastic model, e.g. imprecise moment information or samples. To characterize the partial information about the true distribution, we leverage Wasserstein ambiguity sets \cite{zhao2018data, mohajerin2018data}, which contain a family of probability measures on the measurable space $(\Omega, X)$.  
\vspace{-1mm}
\subsection{Wasserstein ambiguity set}
A Wasserstein ambiguity set is modelled as a Wasserstein ball centered at a discrete empirical distribution. The Wasserstein ball is a discrepancy-based model wherein the distance  between probability distributions on the probability distribution space and the empirical distribution at the ball center is described by the Wasserstein metric. The Wasserstein metric defines the distance between all probability distributions $\mathbb{Q}$ supported on $\mathbb{W}_{\xi}$ with finite $p$-moment $\int_{\mathbb{W}_{\xi}}\|\xi\|^{p} \mathbb{Q}(d \xi)<\infty$. We restrict $p = 1$ in this paper.

\begin{definition}[Wasserstein Metric \cite{mohajerin2018data}]
\label{def:Wassertein_metric}
The \emph{Wasserstein metric} of order $p \ge 1$ is defined as $d_w: \mathcal{M}(\mathbb{W}_{\xi}) \times \mathcal{M}(\mathbb{W}_{\xi}) \rightarrow \mathbb{R}$ for all distribution $\mathbb{Q}_1, \mathbb{Q}_2 \in \mathcal{M}(\mathbb{W}_{\xi})$ and arbitrary norm on $\mathbb{R}^{n_\xi}$:
\begin{equation}
\footnotesize
d_{w}\left(\mathbb{Q}_{1}, \mathbb{Q}_{2}\right):=\inf_{\Pi} \int_{\mathbb{W}_{\xi}^{2}}\left\|\xi_{1}-\xi_{2}\right\|^{p} \Pi\left(\mathrm{d} \xi_{1}, \mathrm{~d} \xi_{2}\right)
\label{eq:wasserstein_metric}
\end{equation}
where $\Pi$ is a joint distribution of $\xi_{1}$ and $\xi_{2}$ with marginals $\mathbb{Q}_{1}$ and $\mathbb{Q}_{2}$ respectively.
\end{definition}

The Wasserstein metric originates from the optimal transportation problem \cite{villani2009optimal}, which studies the most efficient way to allocate one mass of distribution to another. In \eqref{eq:wasserstein_metric}, the Wasserstein distance between the distribution $\mathbb{Q}_1$ and $\mathbb{Q}_2$ can be interpreted as the minimal energy spent on the allocation if the Euclidean norm is selected and $p=2$. In the following, we will regard one distribution as the empirical distribution and the other as one of the unknown distributions which we asses whether to include or not in the ambiguity set. All these unknown distributions, whose distance from the empirical distribution is lower than a certain value in the Wasserstein sense, are included (to  construct) the ambiguity set.

Specifically, we define the ambiguity set $\mathcal{P}_k$ at time $k$ centered at the empirical distribution leveraging the Wasserstein metric
\begin{equation}
\footnotesize
\mathbb{B}_{\varepsilon}\left(\hat{\mathbb{P}}_{k}\right):=\left\{\mathbb{Q} \in \mathcal{M}(\mathbb{W}_{\xi}): d_{w}\left(\hat{\mathbb{P}}_{k}, \mathbb{Q}\right) \leq \varepsilon\right\}
\label{eq:Wasserstein_ball}
\end{equation}
which specifies the Wasserstein ball with radius $\varepsilon>0$ around the discrete empirical probability distribution $\hat{\mathbb{P}}_{k}$. The empirical probability distribution  $\hat{\mathbb{P}}_{k}:=\frac{1}{N_k} \sum_{l=1}^{N_k} \delta_{\hat{\xi}^{(i)}}$ is the mean of $N_k$ Dirac distributions which concentrates unit mass at the disturbance realization $\hat{\xi}^{(i)} \in \mathbb{W}_{\xi}$. We denote the training set of offline collected realizations $\xi$ as $\hat{\Xi}_{N_k}:=\left\{\hat{\xi}^{(i)}\right\}_{i \in \mathbb{N}_1^{N_k}} \subset \mathbb{W}_{\xi}$, which contains $N_k$ observed disturbance realizations.

The radius $\varepsilon$ tunes the size of the Wasserstein ball \eqref{eq:Wasserstein_ball}, which should be large enough to contain the true distribution but not unnecessarily large, to prevent it from including irrelevant distributions and making the problem over-conservative \cite{zhao2018data,rahimian2019distributionally}. Furthermore, as a function of the radius, the solution of this Wasserstein ambiguity based DRO lies between the classical robust optimization and stochastic optimization, i.e. sample average approximation of the discrete empirical distribution \cite{mohajerin2018data}. In this work, the ball radius is treated as an adaptive hyperparameter. The impact of the ball radius is illustrated and discussed in Section \ref{sec:num_exam}.

\begin{figure}[thpb]
  \centering
  \includegraphics[width=0.8\textwidth]{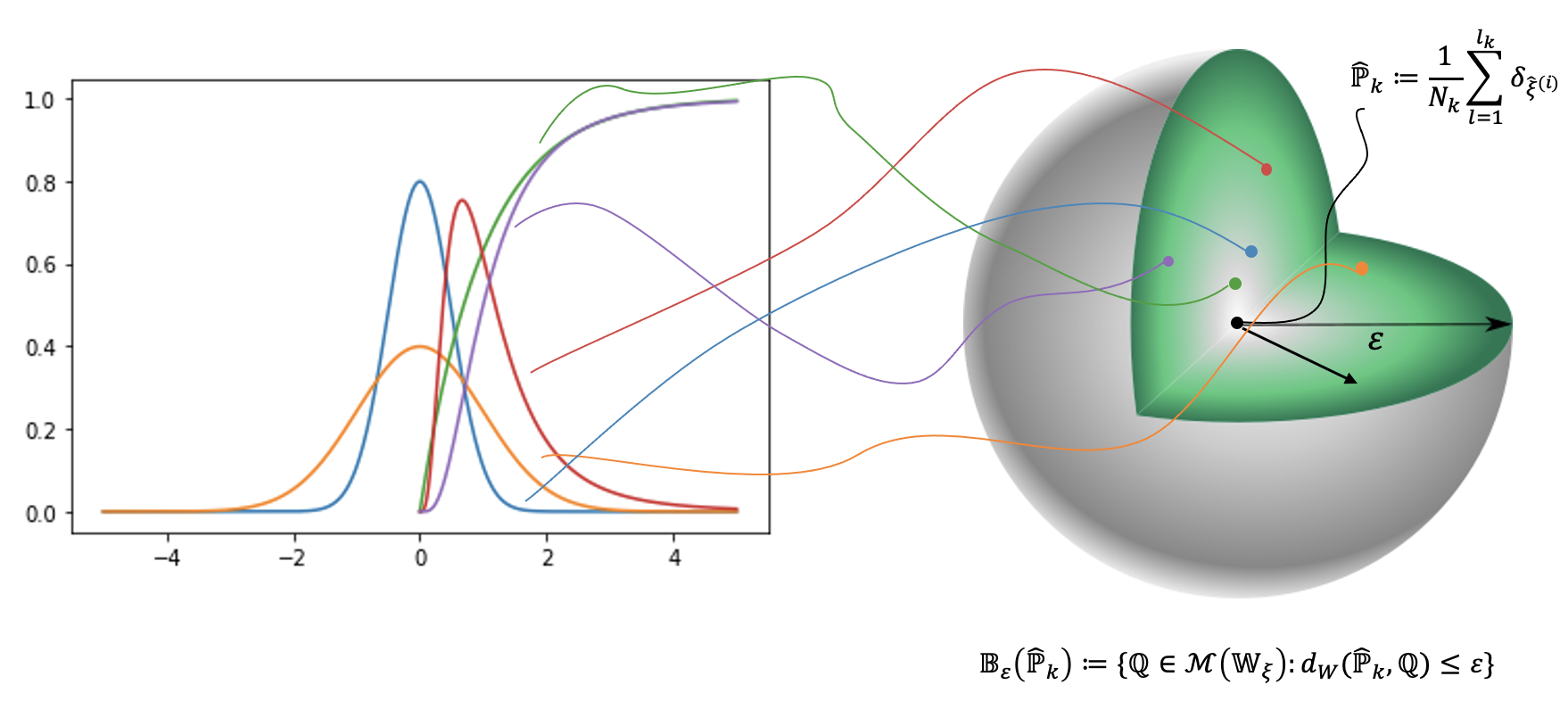}
  \caption{Illustration of Wasserstein ball. Left: multiple probability distributions. Right: the Wasserstein ball containing various probability distributions.}
  \label{fig:wasser_ball}
\end{figure}

\section{Methodology}
The proposed DRMPC scheme \eqref{eq:prototype_prob} is in general not solvable due to the minimax problem over infinite probability distributions and the uncertainty propagation for nonlinear systems. To address the problems mentioned above, we consider two linearization approaches to approximately solve \eqref{eq:prototype_prob}  while guaranteeing the nominal closed-loop state constraint \eqref{eq:nom_state_constraint} can be satisfied with a high probability. The full implementation can be found in the GitHub repository \url{https://github.com/zhengang-zhong/DRMPC_nonlinear_system_via_linearization}.

\subsection{Linearization for Receding Horizon Optimization}
\label{sec:RHO}
In this subsection, we consider two linearization approaches to approximate the original control problem \eqref{eq:prototype_prob}, i.e. 1) successive linearization around nominal states and inputs 2) feedback linearization for nonlinear systems allowing a coordinate transformation. The approximated optimization of \eqref{eq:prototype_prob} is formulated as 
\begin{equation}
\label{eq:approx_prob}
\begin{array}{cl}
\displaystyle\min_{\mathbf{z}, \mathbf{v}} & \mathbb{E}_{\mathbb{P}} \{\sum_{i=0}^{N-1} (\|z_{i\mid k} + e_{i\mid k} + \delta_k \|_{Q}^{2}  + \|v_{i\mid k} + Ke_{i\mid k} \|_{R}^{2}) +  \| z_{N\mid k} + e_{N\mid k}\|_{Q_f}^{2})\} \\
\text { s.t. } 
& z_{0 \mid k}=x_{k}\\
& e_{0 \mid k} \stackrel{a . s .}{=} 0. \\
& z_{i+1 \mid k}=A_k z_{i \mid k}+ B_k v_{i \mid k} + \delta_k   \\
& e_{i+1 \mid k}=A^{cl}_{k}  e_{i \mid k}+D W_{i+k} \\ 
& \displaystyle\sup_{\mathbb{P}_k \in \mathcal{P}_k} \mathbb{E}_{\mathbb{P}_k}\left\{[F]_{j}(z_{i\mid k} + e_{i\mid k} + \delta_k)\right\}  \leq [f]_{j}, \quad k \in \mathbb{N}_{\ge 0}, j \in \mathbb{N}_1^{n_F}, i \in \mathbb{N}_1^N \\
&\displaystyle \max_{e_{i \mid k}} G (K e_{i \mid k}+v_{i \mid k})  \leq g \quad  i \in \mathbb{N}_0^{N-1},\\
\end{array}
\end{equation}
which will be further reformulated into a tractable optimization \eqref{eq:seq_lin_MPC} by leveraging standard linearization techniques as explained in the following subsections.

\subsubsection{Successive linearization}
\label{sec:suc_lin}
Successive linearization, e.g. in \cite{zhakatayev2017successive, cannon2009successive, kuhne2004model}, is an approach approximating the nonlinear dynamics \eqref{eq:system} with a linear time-varying model
\begin{equation}
\label{eq:linearized_system_dist}
x_{i+1\mid k} =A_k x_{i \mid k} + B_k u_{i \mid k}+D W_{i+k} + \delta_k,
\end{equation}
where the matrices $A_{k} \in \mathbb{R}^{n_x \times n_x}$ and $B_{k} \in \mathbb{R}^{n_x \times n_u}$ are the Jacobian matrices of the nonlinear system given by \eqref{eq:system}, and they are defined as follows
\begin{equation}
\label{eq:seq_linear_pair}
\begin{aligned}
A_{k} &=\left.\frac{\partial f\left(x_{i\mid k}, u_{i \mid k}\right)}{\partial x_{i \mid k}}\right|_{x_{i \mid k} = x_{k}, u_{i \mid k} = u_{k-1}} , \\
B_{k} &=\left.\frac{\partial f\left(x_{i\mid k}, u_{i \mid k}\right)}{\partial u_{i \mid k}}\right|_{x_{i \mid k} = x_{k}, u_{i \mid k} = u_{k-1}}, 
\end{aligned}
\end{equation}
where $x_{k}, u_{k-1}$ are current state measurement and input at the previous sampling time, respectively. The linearization error at the sampling time $k$ is defined as $\delta_{k} := f(x_{k}, u_{k-1}) - A_k x_{k} - B_k u_{k-1}$.

\subsubsection{Feedback linearization}
\label{sec:feed_lin}
Similarly to previous work applying feedback linearization to discreet-time MPC \cite{kothare1995robust, simon2013nonlinear}, we first restrict to control-affine nonlinear systems in continuous time of the form:
\begin{equation}
\label{eq:SISO_nonlinear}
\begin{aligned}
\dot{x} &=f(x)+\sum_{j=1}^{n_u} g_{j}(x) u_{j},\\
y &=h(x)
\end{aligned}
\end{equation}
We also define the Lie derivative of the scalar field $h$ along the vector field $f$ as $\sum_{i=1}^{n} \frac{\partial h}{\partial x_{i}} f_{i}(x)$.  Also the relative degree $r$ is defined as the degree that (1) $L_{g} L_{f}^{k} h(x)=0$ for all $x$ around $p$ and $k \in\{0, \ldots, r-2\}$ and (2) $L_{g} L_{f}^{r-1} h(x) \neq 0$, and $L_{f} h(x)$ means the derivative of $h$ along $f$ at $x$.

For the nonlinear system with inputs and outputs in same dimension, i.e. $\operatorname{dim} y=\operatorname{dim} u=n_u$, we could represent the input-output relation through
$$
\left(\begin{array}{c}y_{1}^{\left(r_{1}\right)} \\ \vdots \\ y_{n_u}^{\left(r_{n_u}\right)}\end{array}\right)=\left(\begin{array}{c}L_{f}^{r_{1}} h_{1}(x) \\ \vdots \\ L_{f}^{r_{n_u}} h_{n_u}(x)\end{array}\right)+
\underbrace{\left(\begin{array}{ccc}L_{g_{1}} L_{f}^{r_{1}-1} h_{1}(x) & \cdots & L_{g_{n_u}} L_{f}^{r_{1}-1} h_{1}(x) \\ \vdots & \ddots & \vdots \\ L_{g_{1}} L_{f}^{r_{n_u}-1} h_{n_u}(x) & \cdots & L_{g_{n_u}} L_{f}^{r_{n_u}-1} h_{n_u}(x)\end{array}\right)}_{\Lambda(x)}
\left(\begin{array}{c}u_{1} \\ \vdots \\ u_{n_u}\end{array}\right),
$$
where $\left(r_{1}, \ldots, r_{n_u}\right)$ is the relative vector for the corresponding output vectors.

If $\Lambda(x)$ is nonsingular, then the static feedback policy $u$ can be defined as 
$$u(x)=\Lambda^{-1}(x)\left(-\left[\begin{array}{c}L_{f}^{r_{1}} h_{1}(x) \\ \vdots \\ L_{f}^{r_{n_u}} h_{n_u}(x)\end{array}\right]+\tilde{u}\right)$$ such that the closed-loop system is linear and the nonlinear relation between the inputs and outputs is decoupled. More detailed can be found in classic nonlinear control materials, e.g. \cite[Chapter 5]{isidori1985nonlinear}, \cite[Section 6.5.2]{kwatny2001nonlinear}.

If the system happens to be the special case satisfying Assumption \ref{assump:SISO_and_full_RD}, i.e. SISO system with full relative degree \cite[Theorem 2]{tse1987linear} \cite[Theorem 4.38]{robenack2017nichtlineare}, then, the nonlinear system presents the charming property that allows it to be reformulated into the control canonical form, i.e. exactly input-state linearizable
$$
\begin{aligned}
\dot{z}_{1} &=z_{2} \\
& \vdots \\
\dot{z}_{n-1} &=z_{n} \\
\dot{z}_{n} &=\alpha(z)+\beta(z) u  \\
y &=z_{1}
\end{aligned}
$$
under $z=\Phi(x)$ with $\alpha(z)=\left.L_{f}^{n} h(x)\right|_{x=\Phi^{-1}(z)}$, 
$\beta(z)=\left.L_{g} L_{f}^{n-1} h(x)\right|_{x=\Phi^{-1}(z)}$ and hence they can be presented as linear controllable system in the Brunovsky canonical form
$$
\begin{aligned}
&\dot{z}=Az+B\tilde{u} \\
&y=c^{T} z
\end{aligned}
$$
under the feedback 
\begin{equation}
\label{eq:feedback_input_state}
u=\frac{1}{\beta(z)}(\tilde{u}-\alpha(z)).
\end{equation}

\begin{assumption}
\label{assump:SISO_and_full_RD}
The nonlinear system \eqref{eq:SISO_nonlinear} is single-input-single-output (SISO) and its relative degree $r = n_x$.
\end{assumption}

For the nonlinear system which is exactly input-to-state linearizable, we describe the corresponding discrete-time system as 
$$
z_{k+1}=A z_{k}+B \tilde{u}_{k},
$$ and the system input as $u=\frac{1}{\beta(z_k)}(\tilde{u}_k-\alpha(z_k))$,
similiar to \cite{simon2013nonlinear}. The corresponding formulation for the disturbed system is described as 
\begin{equation}
\label{eq:feedback_sys}
z_{k+1}=A z_{k}+B \tilde{u}_{k} + D w_k. 
\end{equation}

\begin{remark}
This feedback linearization method can only be applied to systems that are exactly input-to-state linearizable. The nonlinear feedback will transform the original polytopic input constraints into non-convex nonlinear constraints. Hence, to maintain a convex program for the MPC problem, convex approximations of the transformed input constraints are considered. Authors in \cite{simon2013nonlinear} recommend replacing nonconvex input constraints with their global inner convex polytopic approximations, local approximations are constructed  dynamically to reduce the conservativeness caused by global approximations.
\end{remark}

\subsection{Reformulation of Predicted Linearized Systems}
We are now at the stage of reformulating the DRMPC \eqref{eq:prototype_prob} into tractable forms for the both linearized systems. To reduced the computational complexity of full disturbance feedback policy in \cite{zhong2021data}, we leverage the tube-based MPC for the tractable reformulation and subsequent algorithm. In the context of tube-based MPC \cite{lorenzen2016constraint,hewing2020recursively, kouvaritakis2010explicit}, the system states within the prediction horizon are separated into two parts $x_{i \mid k}=z_{i \mid k}+e_{i \mid k}$, where the nominal part $z_{i \mid k}=\mathbb{E}_{k}\left\{x_{i \mid k}\right\}$ and the stochastic error part $e_{i \mid k}$ due to the disturbance r.v. sequence $\Xi_{k}$.

From here upon, we denote the linearization pair as $(A_k, B_k)$ without distinguishing whether it comes from \eqref{eq:seq_linear_pair} or \eqref{eq:feedback_sys}, despite the fact that the system matrices are time-varying in \eqref{eq:seq_linear_pair} and remain static in \eqref{eq:feedback_sys}.  Similarly, the linearization error $\delta_k$ takes the value $0$ for \eqref{eq:feedback_sys} or the sequential linearization error around the measurement at time $k$ for \eqref{eq:seq_linear_pair}. We assume that the linearization pair $(A_k,B_k)$ is stabilizable such that we can find a stabilizing feedback gain $K$ guaranteeing $A^{cl}_{k} = A_k +B_kK$ is stable \cite{chen1999linear}. Similar as in \cite{hewing2020recursively}, we define the predicted input with the prestabilization gain $K$ as 
\begin{equation}
\label{eq:control_law}
u_{i \mid k}=K e_{i \mid k}+v_{i \mid k},
\end{equation}
where $v_{i \mid k} \in \mathbb{R}^{n_u}$ are decision variables in the optimal control problem. The prediction of the split two parts can therefore be formulated as
\begin{equation}
\label{eq:tube_form}
\begin{array}{ll}
z_{i+1 \mid k}=A_k z_{i \mid k}+ B_k v_{i \mid k} + \delta_k & z_{0 \mid k}=x_{k} \\
e_{i+1 \mid k}=A^{cl}_{k} e_{i \mid k}+D W_{i+k} & e_{0 \mid k} \stackrel{a . s .}{=} 0. 
\end{array}
\end{equation} The purpose of selecting a stabilizing feedback gain $K$ is to shrink the closed-loop propagation of the predicted error such that the input is less conservative.

After introducing tube-based MPC, the objective function \eqref{eq:proto_obj} at time $k$ is $ \mathbb{E}_{\mathbb{P}} \{\sum_{i=0}^{N-1} (\|z_{i\mid k} + e_{i\mid k}  \|_{Q}  + \|v_{i\mid k} + Ke_{i\mid k} \|_{R}) +  \| z_{N\mid k} + e_{N\mid k}\|_{Q_f})\}$; similar reformulations for the state and input constraints \eqref{eq:proto_constraint} can be acquired:
\begin{equation}
\label{eq:state_constraints}
\displaystyle\sup_{\mathbb{P}_k \in \mathcal{P}_k} \mathbb{E}_{\mathbb{P}_k}\left\{[F]_{j}(z_{i\mid k} + e_{i\mid k})\right\}  \leq [f]_{j}, \quad k \in \mathbb{N}_{\ge 0}, j \in \mathbb{N}_1^{n_F}, i \in \mathbb{N}_1^N 
\end{equation}
 and $$\max_{e_{i \mid k}} G (K e_{i \mid k}+v_{i \mid k})  \leq g \, \quad  i \in \mathbb{N}_0^{N-1}.$$

After solving the optimal control problem \eqref{prob:orig_prob}, the input applied to the closed-loop system is the first element of the decision sequence $\boldsymbol{v}$, namely
\begin{equation}
\label{eq:opt_input}
u_k = v^{*}_{0\mid k}
\end{equation}
for \eqref{eq:seq_linear_pair} and 
\begin{equation}
u_k=\frac{1}{\beta(x_k)}(v^{*}_{0\mid k}-\alpha(x_k))
\end{equation}
for \eqref{eq:feedback_sys}.

After introducing the tube-based MPC formulation, we now describe the tractable conic reformulation for the DRMPC \eqref{eq:prototype_prob}. To reformulate the objective function, we introduce a further assumption on the disturbance expectation and variance.

\begin{assumption}[Zero expected disturbance]
\label{asp:iid}
We assume that in the discrete-time nonlinear system \eqref{eq:system}, the disturbance $w_k$ is an i.i.d. random process with the zero expectation and an unknown but fixed variance $\sigma_k$ for all $k \in \mathbb{N}_{\ge 0}$.
\end{assumption}

The i.i.d. random process is a common assumption made in control literature, e.g. \cite{arcari2020dual, coppens2021data}. It assumes a priori that only the first  moment of the random process is acquired as an extra partial distributional information, which can either be estimated or prescribed a priori \cite{wan2014estimating}. We assume for simplicity that the expectation is zero; however, this can be extended to any expectation. Also, the variance is required to be static but unknown; the formulation below is not dependent on the variance value.

\begin{theorem}
\label{th:exact_reform}
Suppose that Assumptions \ref{assump:iid},\ref{assump:SISO_and_full_RD},\ref{asp:iid} hold, then the distributionally robust chance constrained optimization problem \eqref{eq:approx_prob} is equal to the following tractable convex optimization problem
\begin{equation}
\label{eq:seq_lin_MPC}
\begin{array}{cl}
\displaystyle\min_{
\substack{\boldsymbol{z}, \boldsymbol{v}\\
    \gamma \ge 0, \lambda \ge 0, s, \xi_{dual1}\ge 0}}
      & \sum_{i=1}^{N-1} z_{i\mid k}^{\top} Q z_{i\mid k}+v_{i\mid k}^{\top} R v_{i\mid k}+z_{N\mid k}^{\top} Q z_{N \mid k} \\
\text { s.t. } 
& z_{0\mid k}=x_{k}, \\
& z_{i+1\mid k}=A z_{i\mid k}+B v_{i\mid k} + \delta_k\\
\begin{aligned}
i & \in \mathbb{N}_0^{N-1}\\  j & \in \mathbb{N}_1^{n_G}
\end{aligned}
&
\left\{\begin{aligned}
&h^{\top} \xi_{\text {dual } 1, i j} \leq I_{j}^{\top}\left(g-G v_{i}\right)\\
&H^{\top} \xi_{dual1,ij} = L_{ij}\\
& \xi_{dual1,ij} \ge 0
\end{aligned}\right.\\
\begin{aligned}
i & \in \mathbb{N}_1^{N-1}\\  j & \in \mathbb{N}_1^{n_F}
\end{aligned}
&
\left\{\begin{aligned}
& \lambda_{ij} \varepsilon + \frac{1}{N_k} \sum_{l=1}^{N_k}s_{ijl}  \le 0\\
& I_j^{\top}(F (z_{i} +[D_x]_{i \times n_x + 1: (i+1) \times n_x }\hat{\xi}^{(l)}) - f )+ \gamma_{ijl}^{\top}(h_{\xi}-H_{\xi} \hat{\xi}^{(l)})  \le s_{ijl}\\
& \|H_{\xi}^{\top}\gamma_{ijl} -  I_j^{\top}(F[D_x]_{i \times n_x + 1: (i+1) \times n_x}) \|_{\infty} \le \lambda_{ij}\\
& \gamma_{ijl}  \ge 0\\
\end{aligned}\right.\\
\begin{aligned}
i &= N\\  j & \in \mathbb{N}_1^{n_{F_N}}
\end{aligned}
&
\left\{
\begin{aligned}
&\lambda_{ij} \varepsilon + \frac{1}{N_k} \sum_{l=1}^{N_k}s_{ijl}  \le 0\\
& I_j^{\top}(F_N (z_{i}+[D_x]_{i \times n_x + 1: (i+1) \times n_x }\hat{\xi}^{(l)}) - f_N)+ \gamma_{ijl}^{\top} (h_{\xi}-H_{\xi} \hat{\xi}^{(l)} ) \le s_{ijl}\\
& \|H_{\xi}^{\top}\gamma_{ijl} -  I_j^{\top}(F_N[D_x]_{i \times n_x + 1: (i+1) \times n_x }) \|_{\infty}  \le \lambda_{ij}\\
& \gamma_{ijl}  \ge 0
\end{aligned}\right. \\
& l \in \mathbb{N}_1^{N_k},
\end{array}
\end{equation}
where $D_{u}=\left[\begin{array}{ccccc}
0   & \ldots & 0 & 0\\
K(A_{k}^{cl})^{0}D  & \ldots & 0 & 0\\
\vdots  & \ddots  & 0 & 0\\
K(A_{k}^{cl})^{N-1}D   & \ldots & K(A_{k}^{cl})^{0}D & 0
\end{array}\right]
$, $D_{x}=\left[\begin{array}{ccccc}
0   & \ldots & 0 \\
(A_{k}^{cl})^{0}D  & \ldots & 0 \\
\vdots  & \ddots  & 0 \\
(A_{k}^{cl})^{N-1}D   & \ldots & (A_{k}^{cl})^{0}D 
\end{array}\right]
$, $L_{ij}^{\top} = I^{\top}_j G[D_u]_{i \times n_u + 1: (i+1) \times n_u }$ and $I_i$ is a column vector, in which only the i-th entry is $1$ and the remaining entries are $0$.
\end{theorem}

\begin{proof}
We first introduce the formulation of the predicted state under control laws \eqref{eq:control_law} within the prediction horizon $N$. Under Assumption \ref{asp:iid} and a linear approximation of the system, the expectation of each predicted error $e_{i \mid k}$ is zero. Hence, the expected quadratic objective function can be reformulated to $\sum_{i=1}^{N-1} z_{i\mid k}^{\top} Q z_{i\mid k}+v_{i\mid k}^{\top} R v_{i\mid k}+z_{N\mid k}^{\top} Q z_{N \mid k} + c$, where $c$ is a constant \cite{lorenzen2016constraint}.

Then, we reformulate the feasibility set for worst-case input constraints $\{v_{i \mid k} \mid \exists v_{i \mid k}  s.t. \max_{e_{i \mid k}} G (K e_{i \mid k}+v_{i \mid k})  \leq g \quad  i \in \mathbb{N}_0^{N-1} \}$  into feasibility sets of linear constraints. Given that the accumulated error vector $e_{i+1 \mid k} = A^{cl}_{k}  e_{i \mid k}+D W_{i+k}$ is linear in unknown disturbances prior to $i+1$ within the prediction, we denote $e_{i+1 \mid k} = [D_x]_{(i + 1) \times n_x + 1 : (i+2)\times n_x} \Xi_{k}$.
Consider the dual problem of each separate left-hand-side equation $\max_{e_{i \mid k}} I^{\top}_{j}G (K e_{i \mid k}+v_{i \mid k})$, we acquire minimization problems \cite{sun2021robust}
\begin{equation}
\label{eq:input_reform1}
\begin{aligned}
\min_{\xi_{\text {dual } 1, i j}} & h_{\xi}^{\top} \xi_{\text {dual } 1, i j} + I_{j}^{\top}\left(G v_{i \mid k}\right)\\
\text{s.t.} & H_{\xi}^{\top} \xi_{dual1,ij} = L_{ij}\\
& \xi_{dual1,ij} \ge 0,
\end{aligned}
\end{equation}
where $L_{ij}^{\top} = I^{\top}_j G[D_u]_{i \times n_u + 1: (i+1) \times n_u }$.
As additive disturbances are bounded, their accumulated errors $e_{i \mid k}$ within the prediction horizon are also bounded. Hence the left-hand-side equation $\max_{e_{i \mid k}} G (K e_{i \mid k}+v_{i \mid k})$ has a finite optimal value for any given $v_{i \mid k}$, so as its dual  \cite[Proposition 5.2.1]{bertsekas2009convex}. Based on the duality, the worst-case input constraints can be reformulated as 
\begin{equation}
\label{eq:input_reform2}
\{I^{\top}_j v_{i \mid k} \mid \exists v_{i \mid k}, \xi_{\text {dual } 1, i j}  \quad  \text{s.t.} h^{\top} \xi_{\text {dual } 1, i j} \leq I_{j}^{\top}\left(g-G v_{i}\right), H_{\xi}^{\top} \xi_{dual1,ij} = L_{ij}
 \xi_{dual1,ij} \ge 0,\quad  i \in \mathbb{N}_0^{N-1}, j \in \mathbb{N}_{1}^{n_G} \}\end{equation}.
 
Finally, we reformulate the feasibility set for the state constraints defined by the data-driven distributionally robust optimization using the Wasserstein ambiguity in \eqref{eq:state_constraints} into feasibility sets of linear constraints. The feasibility set is equivalently defined as \begin{equation} \{z_{i \mid k}  \mid \exists z_{i \mid k} s.t.
\displaystyle\sup_{\mathbb{P}_k \in \mathcal{P}_k} \mathbb{E}_{\mathbb{P}_k}\left\{[F]_{j}(z_{i\mid k} + e_{i\mid k}) - [f]_{j}\right\}  \leq 0, \quad k \in \mathbb{N}_{\ge 0}, j \in \mathbb{N}_1^{n_F}, i \in \mathbb{N}_1^N  \}
\end{equation}
Similar to input constraints, the left-hand-side equation $I^{\top}_{j}F_{j}(z_{i\mid k} + e_{i\mid k} ) - [f]_{j}$ is affine in uncertainty $\Xi_k$. Each equation can be described as $I^{\top}_{j}F(z_{i\mid k} + [D_x]_{i \times n_x + 1 : (i+1)\times n_x} \Xi_{k} ) - [f]_{j}$. As all samples of the disturbance sequence realizations are drawn following the i.i.d. assumption, the empirical distribution as the center of the Wasserstein ball is formulated as $\sum_{l = 1}^{N_k} \hat{\xi}^{(l)}$, where $N_k$ denotes the number of samples applied the instance $k$. According to \cite{mohajerin2018data, zhong2021data}, $\displaystyle\sup_{\mathbb{P}_k \in \mathcal{P}_k} \mathbb{E}_{\mathbb{P}_k}\left\{[F]_{j}(z_{i\mid k} + e_{i\mid k} ) - [f]_{j}\right\}$ is equal to 
\begin{equation}
\label{eq:state_constraints_reform}
\begin{aligned}
&
\begin{drcases}
\min_{\lambda_{ij}, \gamma_{ijl}, z_{i\mid k}} \lambda_{ij} \varepsilon + \frac{1}{N_k} \sum_{l=1}^{N_k}s_{ijl} \\
I_j^{\top}(F (z_{i\mid k} +[D_x]_{i \times n_x + 1: (i+1) \times n_x}\hat{\xi}^{(l)}) - f)+ \gamma_{ijl}^{\top}(h_{\xi}-H_{\xi} \hat{\xi}^{(l)})  \le s_{ijl}\\
\|H_{\xi}^{\top}\gamma_{ijl} -  I_j^{\top}(F[D_x]_{i \times n_x + 1: (i+1) \times n_x}) \|_{\infty}  \le \lambda_{ij}\\
\gamma_{ijl}  \ge 0\\
\end{drcases}
\begin{aligned}
i  &\in \mathbb{N}_1^{N-1} \\  j &\in \mathbb{N}_1^{n_{F}}\\
\end{aligned}\\
& \forall l \in \mathbb{N}_1^{N_k}.
\end{aligned}
\end{equation}
Hence, the feasibility set is described as \begin{equation}
\label{eq:state_feasible}
\{z_{i \mid k}  \mid \exists \lambda_{ij}, \gamma_{ijl}, z_{i \mid k} s.t.
\begin{aligned}
&
\begin{drcases}
\lambda_{ij} \varepsilon + \frac{1}{N_k} \sum_{l=1}^{N_k}s_{ijl} \le 0\\
I_j^{\top}(F (z_{i \mid k} +[D_x]_{i \times n_x + 1: (i+1) \times n_x}\hat{\xi}^{(l)}) - f)+ \gamma_{ijl}^{\top}(h_{\xi}-H_{\xi} \hat{\xi}^{(l)})  \le s_{ijl}\\
\|H_{\xi}^{\top}\gamma_{ijl} -  I_j^{\top}(F[D_x]_{i \times n_x + 1: (i+1) \times n_x}) \|_{\infty}  \le \lambda_{ij}\\
\gamma_{ijl}  \ge 0\\
\end{drcases}
\begin{aligned}
i  &\in \mathbb{N}_1^{N-1} \\  j &\in \mathbb{N}_1^{n_{F}}\\
\end{aligned}\\
& \forall l \in \mathbb{N}_1^{N_k}.
\end{aligned}
\end{equation}
Similarly, we could get the feasibility set of terminal states.

The proof is complete if the optimization problem \eqref{eq:seq_lin_MPC} has a non-empty solution, resulting in an non-empty feasibility set of \eqref{eq:input_reform2} and \eqref{eq:state_feasible},  which means that the constraints in \eqref{eq:approx_prob} are satisfied. Also both optimization problems share the same optimal objective function.
\end{proof}

The main ideas behind this reformulation is transforming the original infinite-dimensional DRO into a finite-dimensional convex optimization. Also, the structure imposed by the Wasserstein ball after reformulation plays a similar role of constraint backoffs in classic stochastic/robust MPC. The nominal state constraints are hence tightened with respect to the worst-case distribution within the ambiguity set of the empirical distribution. From the reformulation \eqref{eq:state_feasible} it can be observed that each distributionally robust state constraint imposed for all states within the prediction horizon will result in $4\times (N - 1) \times N_k \times n_F  $ extra constraints, and $(N - 1) \times (n_F + n_F \times N_k  + n_F \times N_k \times n_h)$ extra auxiliary decision variables, where $n_h$ is the number of polytopic constraints of additive disturbances. This is the price to pay to be distributionally robust, however, given that this is a convex program, its solution is still efficient.

\subsection{Reformulation for tightened state constraint}
Note that if we replace the distributionally robust state constraint in \eqref{eq:prototype_prob} by slightly more conservative constraints 
\begin{equation}
\label{eq:new_state_constraints}
\displaystyle\max_{\mathbb{P}_k \in \mathcal{P}_k} \mathbb{E}_{\mathbb{P}_k}\left\{ \max_{i} I^{\top}_i (\tilde{F} x_{[1,N] \mid k} - \tilde{f})\right\}  \leq 0, \quad k \in \mathbb{N}_{ \ge 0}, i \in \mathbb{N}_1^{(N-1) \times n_F + n_{F_N}}
\end{equation}
 requiring that the point-wise maximum of state violation is bounded in the distributionally robust sense. Here we provide the stacked version of $F$, $f$ and $x_{i,x}$ $\tilde{F} \triangleq \begin{bmatrix} 
\smash{\underbrace{\begin{matrix}
        F&\dotsb&F
    \end{matrix}}_{N-1}} & F_N
\end{bmatrix}^{\top}$, $\tilde{f} \triangleq \begin{bmatrix} 
\smash{\underbrace{\begin{matrix}
        f&\dotsb&f
    \end{matrix}}_{N-1}} & f_N
\end{bmatrix}^{\top}$ and $x_{[1,N] \mid k} \triangleq \begin{bmatrix}
x_{1 \mid k}^{\top}, \dotsb, x_{N \mid k}^{\top}
\end{bmatrix}^{\top}$.

The constraints in \eqref{eq:new_state_constraints} mean that the maximal expected violations over all $(N-1) \times n_F + n_{F_N}$ state constraints is upper bounded by zero, where the expectation is taken with respect to the worst-case distribution within the ambiguity set. By leveraging similar technique used for the state constraints reformulation in \eqref{eq:state_constraints_reform}, we will get state constraints in a compacter form. Also with the same objective function and input constraints as in \eqref{eq:seq_lin_MPC}, we get the corresponding conic optimization problem for the newly imposed state constraints \eqref{eq:new_state_constraints} as follows
\begin{equation}
\label{eq:seq_lin_MPC_new}
\begin{array}{cl}
\displaystyle\min_{
\substack{\boldsymbol{z}, \boldsymbol{v}\\
    \gamma \ge 0, \lambda \ge 0, s, \xi_{dual1}\ge 0}}
      & \sum_{i=1}^{N-1} z_{i\mid k}^{\top} Q z_{i\mid k}+v_{i\mid k}^{\top} R v_{i\mid k}+z_{N\mid k}^{\top} Q z_{N \mid k} \\
\text { s.t. } 
& z_{0\mid k}=x_{k}, \\
& z_{i+1\mid k}=A z_{i\mid k}+B v_{i\mid k} + \delta_k\\
\begin{aligned}
i & \in \mathbb{N}_0^{N-1}\\  j & \in \mathbb{N}_1^{n_G}
\end{aligned}

&
\left\{\begin{aligned}
&h^{\top} \xi_{\text {dual } 1, i j} \leq I_{j}^{\top}\left(g-G v_{i}\right)\\
&H^{\top} \xi_{dual1,ij} = L_{ij}\\
& \xi_{dual1,ij} \ge 0
\end{aligned}\right.\\
\begin{aligned}
j & \in \mathbb{N}_1^{(N-1) \times n_F + n_{F_N}}
\end{aligned}
&
\left\{\begin{aligned}
& \lambda_{j} \varepsilon + \frac{1}{N_k} \sum_{l=1}^{N_k}s_{jl}  \le 0\\
& I_j^{\top}(\tilde{F} (\mathbf{z} +D_x\hat{\xi}^{(l)}) - \tilde{f} )+ \gamma_{jl}^{\top}(h_{\xi}-H_{\xi} \hat{\xi}^{(l)})  \le s_{jl}\\
& \|H_{\xi}^{\top}\gamma_{jl} -  I_j^{\top}(\tilde{F}D_x) \|_{\infty} \le \lambda_{j}\\
& \gamma_{jl}  \ge 0\\
\end{aligned}\right.\\
& \forall l \in \mathbb{N}_1^{N_k},
\end{array}
\end{equation}

\begin{figure}[thpb]
  \centering
  \includegraphics[height = 0.6\textheight, width=0.6\textwidth]{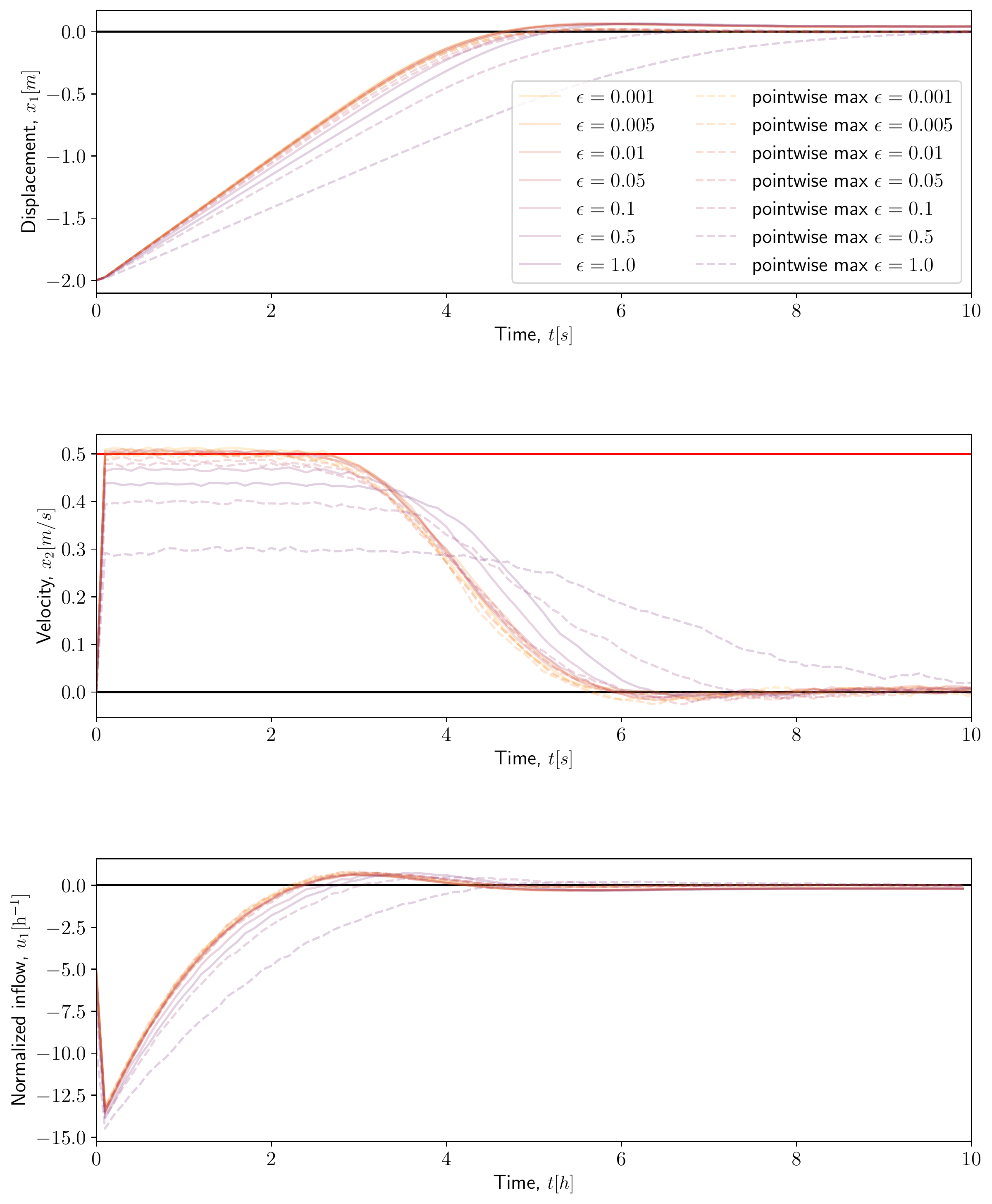}
  \caption{Illustration of the comparison between two conic formulations applying successive linearization. The solid line is the expected state trajectory of \eqref{eq:seq_lin_MPC} and the dashed line is the expected state trajectory of \eqref{eq:seq_lin_MPC_new}}
  \label{fig:two_conic}
\end{figure}

This formulation of state constraints will result in a more conservative system behavior than in \eqref{eq:seq_lin_MPC}, which can be interpreted with the aid of the Jensen's inequality when the order of operations $\mathbb{E}_{\mathbb{P}_{k}}$ and $\max_{i}$ are exchanged. This statement can be verified through the simulation result in the section \ref{sec:num_exam} via Fig \ref{fig:two_conic}.




\subsection{Algorithm}
The proposed DRMPC in this paper has only one extra tuning parameter than a classic MPC approaches. This parameter is the Wasserstein ball radius. In general, if the radius is set too small, it is unlikely for the ball to contain the true distribution when the sample number is very small. This ball is centered at the empirical distribution, and hence,  if the number of samples is very small, the method is likely to result in state constraint violations given a bad estimation of the ambiguity set. To mitigate this, as time progresses, this algorithm will collect more samples during the control process and increase the rate of state constraint satisfaction. The algorithm starts with $N_s$ samples of disturbances collected offline and contained in the data set $\mathcal{D}$. At each sampling time $k$, the conic optimization problem for DRMPC is solved with $N_s$ samples. After solving the optimization problem, a control action is determined based on the first element of the optimal control sequence. The additive disturbance at the sampling time $k$ will be collected and after every $N$ sampling times a newly collected sample will be included in the data set $\mathcal{D}$ until the collected number $N_c$ achieved the prescribed bound of sample numbers $N_{cmax}$. The algorithm is summarized as below.

 \begin{algorithm}[H]
 \caption{Distributionally robust MPC}
 \begin{algorithmic}[1]
  \label{alg:1}
   \State \textbf{Input: $A$, $B$, $D$, $Q$, $R$, $K$, $F$, $f$, $G$, $g$, $\varepsilon$, $H_\xi$, $h_\xi$, $\alpha$, $\beta$}
    \State \textbf{Initialize} $x_0$, $\mathcal{D}$, $N_s$, $N$, $k=0$, $N_c = 0$, $N_{cmax}$, $d_\xi = \varnothing$
    \While{True}
    
  \If{($k$ mod $N$ == $0$ and $k > 0$ and $N_c < N_{cmax}$ )}
  \State Append new data $d_{\xi}$ into $\mathcal{D}$ and let $d_\xi = \varnothing$.
  \State $N_s = N_s + 1$.
  \EndIf
    \State Acquire current state $x_k$.
    \State Select $N_s$ samples from $\mathcal{D}$ to formulate $\widehat{\Xi}_{N_s}$.
  \State Solve \eqref{eq:seq_lin_MPC} or \eqref{eq:seq_lin_MPC_new} to acquire the optimal input sequence $\mathbf{v}_{k}^{*}$.
  \State $u_k = v^{*}_{0\mid k}$ if sequential linearization applied or $u_k=\frac{1}{\beta(x_k)}(v^{*}_{0\mid k}-\alpha(x_k))$ if feedback linearization applied.
    \State Acquire the additive disturbance $w_k$ and append in the data sequence $d_\xi$.
  \State k = k+1
    \EndWhile
 \end{algorithmic} 
 \end{algorithm}

\section{Case Study}
Two dynamic processes are used as case studies to illustrate DRMPC's properties. As the focus is on a clear illustration, we qualitatively compare different methods on a conceptual example and demonstrate the capability of our framework in a complex reactor process. In both case studies, we place the emphasis on constraints satisfaction while visualizing the tracking task can be realized successfully.
\label{sec:num_exam}

\subsection{Nonlinear mass spring system}
The first system considered is a nonlinear mass spring system adapted from  \cite{chen1984linear} with $m = 2\,\text{kg}, k_1 = 3\, \text{N/m}, k_2 = 2 \,\text{N/m}$:
$$
\begin{aligned}
\dot{x}_{1} &= x_{2}\\
\dot{x}_{2} &= -\frac{k_{2}}{m} x_{1}^{3}-\frac{k_{1}}{m} x_{2}+\frac{1}{m} u
\end{aligned}
$$
The discrete-time system is acquired by using the Runge-Kutta method with fourth order with the sampling period $0.1\,\text{s}$. We simulate the control performance for the discrete-time system suffering from the additive disturbance bounded within $[-1, 1]$ on the state element $x_2$. Two types of distributions are considered here: (1) $\sin(W)$, where $W \sim \mathcal{N}\left(0, 1\right)$; (2) uniform distribution. The prediction horizon for this system is set to $N = 5$.

The control goal of this system is to track the state $x_r = [0, 0]$ starting from the initial state $x_{init} = [-2,0]$, while satisfying the state constraint $x_2 \le 0.5\,\text{m/s}$. The parameters are selected as $Q = Q_f =  \begin{bmatrix}
100 & 0 \\
0 & 1
\end{bmatrix}$, $R = [1]$, $K = \begin{bmatrix}
-9.034 & -3.850
\end{bmatrix}^{\top}$.

\subsection{CSTR}
The second case study considers the CSTR control problem from \cite{klatt1998gain}. Within such a CSTR, cyclopentenol is produced from cyclopentadiene by acid-catalyzed electrophylic hydration in aqueous solution \cite{chen1995nonlinear}. Such a process can be formulated as the following nonlinear differential equations
\begin{equation}
\label{eq:CSTR_ODE}
\begin{aligned}
\frac{d c_{\mathrm{A}}}{d t} &=\frac{\dot{V}}{V_{\mathrm{R}}}\left(c_{\mathrm{A} 0}-c_{\mathrm{A}}\right)-k_{1} c_{\mathrm{A}}-k_{3} c_{\mathrm{A}}^{2}+w_{1} \\
\frac{d c_{\mathrm{B}}}{d t} &=-\frac{\dot{V}}{V_{\mathrm{R}}} c_{\mathrm{B}}+k_{1} c_{\mathrm{A}}-k_{2} c_{\mathrm{B}} \\
\frac{d \vartheta}{d t}=& \frac{\dot{V}}{V_{\mathrm{R}}}\left(\vartheta_{0}-\vartheta\right)+\frac{k_{\mathrm{W}} A_{\mathrm{R}}}{\varrho C_{\mathrm{p}} V_{\mathrm{R}}}\left(\vartheta_{\mathrm{K}}-\vartheta\right) \\
&-\frac{k_{1} c_{\mathrm{A}} \Delta H_{\mathrm{R}}^{\mathrm{AB}}+k_{2} c_{\mathrm{B}} \Delta H_{\mathrm{R}}^{\mathrm{BC}}+k_{3} c_{\mathrm{A}}^{2} \Delta H_{\mathrm{R}}^{\mathrm{AD}}}{\varrho C_{\mathrm{p}}}\\
\frac{d \vartheta_{\mathrm{K}}}{d t} &=\frac{1}{m_{\mathrm{K}} C_{\mathrm{pK}}}\left[\dot{Q}_{\mathrm{K}}+k_{\mathrm{W}} A_{\mathrm{R}}\left(\vartheta-\vartheta_{\mathrm{K}}\right)\right]+w_{2} ,
\end{aligned}
\end{equation}
where $k_{i}(\vartheta)=k_{0 i} \cdot \exp \left(\frac{-E_{i}}{R(\vartheta+273.15)}\right)$.   The manipulated variables are $\frac{\dot{V}}{V_R}$ and $\dot{Q}_K$, which represent the inflow
normalized by the reactor volume and the amount of
heat removed by the coolant, respectively.
The corresponding parameters of this CSTR system are summarized within the following table
$$
\begin{array}{|c|c|c|}
\hline \text { Name of parameter } & \text { Symbol } & \text { Value of parameter } \\
\hline \text { collision factor for reaction } k_{1} & k_{01} & 1.287 \cdot 10^{12} \mathrm{~h}^{-1} \\
\text { collision factor for reaction } k_{2} & k_{02} & 1.287 \cdot 10^{12} \mathrm{~h}^{-1} \\
\text { collision factor for reaction } k_{3} & k_{03} & 9.043 \cdot 10^{9} \frac{1}{\mathrm{molA} \cdot h} \\
\text { activation energy for reaction } k_{1} & E_{1} & 9758.3 \mathrm{~K} \\
\text { activation energy for reaction } k_{2} & E_{2} & 9758.3 \mathrm{~K} \\
\text { activation energy for reaction } k_{3} & E_{3} & 8560 \mathrm{~K} \\
\text { enthalpies of reaction } k_{1} & \Delta H^{R}_{A B} & 4.2\, \frac{\mathrm{kJ}}{\mathrm{molA}} \\
\text { enthalpies of reaction } k_{2} & \Delta H^{R}_{B C} & -11.0\, \frac{\mathrm{k} \cdot \mathrm{J}}{\mathrm{molB}} \\
\text { enthalpies of reaction } k_{3} & \Delta H^{R}_{A D} & -41.85\, \frac{\mathrm{kJ}}{\mathrm{molA}} \\
\text { density } & \rho & 0.9342\, \frac{\mathrm{kg}}{1} \\
\text { heat capacity } & C_{p} & 3.01\, \frac{\mathrm{kJ}}{\mathrm{kg} \cdot \mathrm{K}} \\
\text { heat capacity } & C_{pK} & 2.0\, \frac{\mathrm{kJ}}{\mathrm{kg} \cdot \mathrm{K}} \\
\text { heat transfer coefficient for cooling jacket } & k_{w} & 4032\, \frac{\mathrm{kJ}}{\mathrm{h} \cdot \mathrm{m}^{2} \cdot \mathrm{K}} \\
\text { surface of cooling jacket } & A_{R} & 0.215 \mathrm{~m}{ }^{2} \\
\text { reactor volume } & V_{R} & 10.01 \mathrm{~m} \\
\text { coolant mass } & m_{K} & 5.0 \mathrm{~kg} \\
\hline
\end{array}
$$

$$
\begin{array}{|c|c|}
\hline \text { Main operating point }  & \text { Value of parameter } \\
\hline  c_{A\mid S} & 1.2345\, \mathrm{mol}^{-1} \\
\hline  c_{B\mid S} & 0.9\, \mathrm{mol}^{-1} \\
\hline  \vartheta_{\mid S} &  134.15\, ^{\circ} \mathrm{C} \\
\hline  \vartheta_{K\mid S} &  128.97\, ^{\circ} \mathrm{C} \\
\hline  F_{\mid S} &  18.83\, \mathrm{h}^{-1} \\
\hline  \dot{Q}_{K\mid S} &  -4495.7\, \mathrm{kJ}/\mathrm{h} \\
\hline  c_{A0 \mid S} &  5.1\, \mathrm{mol}/\mathrm{l} \\
\hline
\end{array}
$$

The discrete-time system is acquired by using the Runge-Kutta method with fourth order. We simulate the control performance for the discrete-time system suffering from the additive disturbance bounded within $[-0.001, 0.001]$ and $[-0.1, 0.1]$ on the state element $x_1$ and $x_4$, respectively. Two types of distributions are considered here: (1) $\sin(W)$, where $W \sim \mathcal{N}\left(0, 1\right)$; (2) uniform distribution.

Similar to the work \cite{chen1995nonlinear}, the control goal is to control the system element $x_2$. We define the control goal in this paper as tracking the target state $x_r = [1.2345, 0.9, 134.15, 128.97]^{\top}$ starting from the initial state $x_{init} = [1.2345, 1.0, 134.15, 128.97]$, while satisfying the state constraint $x_1 \ge 1.233\,\text{m/s}$. The parameters are selected as $Q = Q_f =  \begin{bmatrix}
1 & 0 & 0 & 0 \\
0 & 1 & 0 & 0 \\
0 & 0 & 1 & 0 \\
0 & 0 & 0 & 1
\end{bmatrix}$, $R = 
\begin{bmatrix}
0.1 & 0\\
0 & 0.1\\
\end{bmatrix}$, $K = \begin{bmatrix}
-5.7164, -4.3252, -1.2812, -0.5330\\
-0.0431, -0.0363, -0.0157, -0.0087
\end{bmatrix}^{\top}$.

\section{Results}
\subsection{Comparison between feedback linearization and successive linearization for case study 1}
\label{sec:case_study1_sim1}
\begin{figure}[thpb]
  \begin{minipage}[b]{0.5\textwidth}
  \centering
    \includegraphics[width=\textwidth]{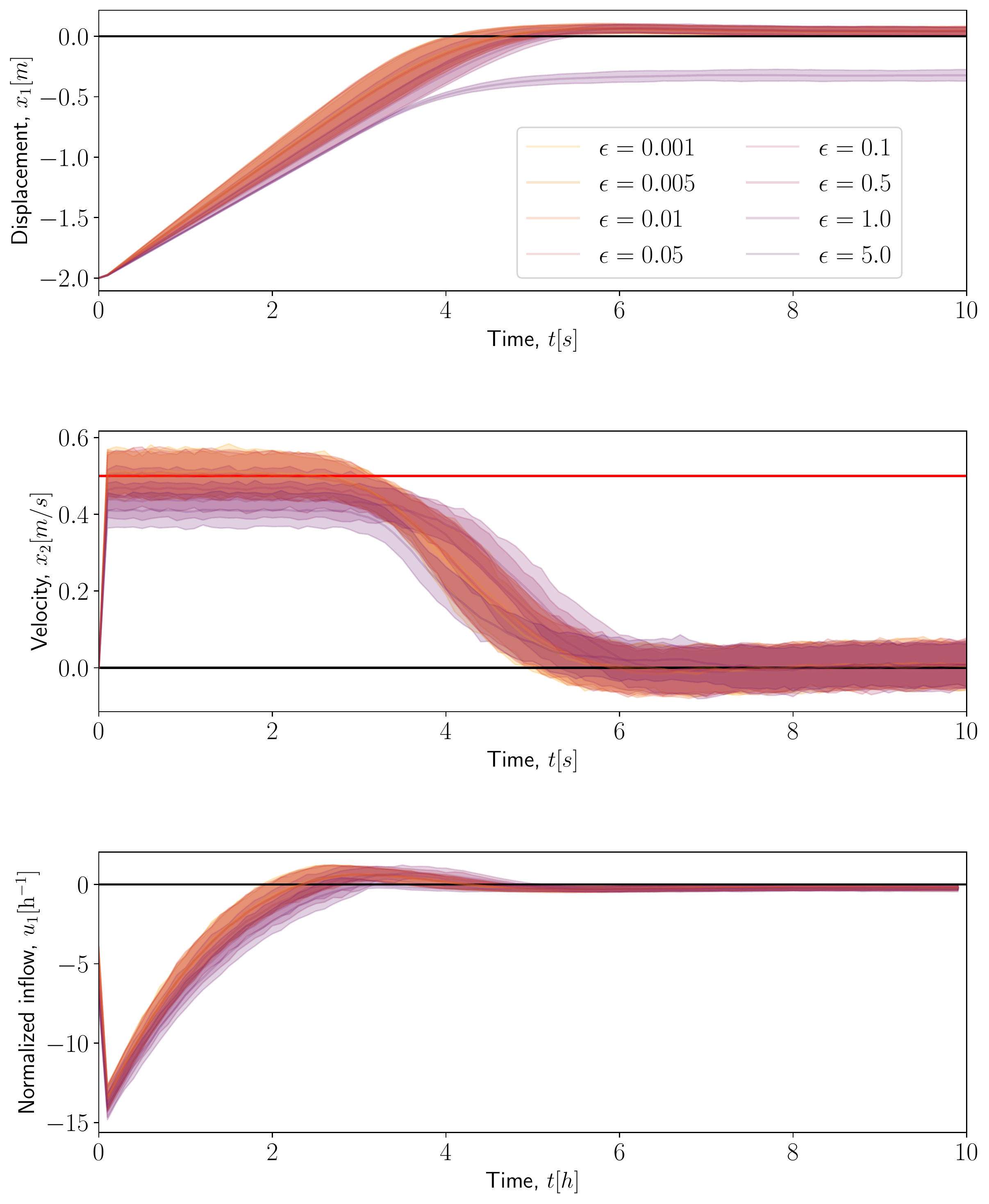}
    \vspace*{-6mm}
    \caption{Simulation results of DRMPC using successive linearization averaged from 500 realizations with one sample and ball radius ranging from $0.001$ to $5$ on the nonlinear mass spring system. Solid lines are the expected trajectories and  shaded areas represent $15-75\%$ percentile of trajectories.}
    \label{fig:academic_sin_suclin}
  \end{minipage}
  \begin{minipage}[b]{0.5\textwidth}
  \centering
    \includegraphics[width=\textwidth]{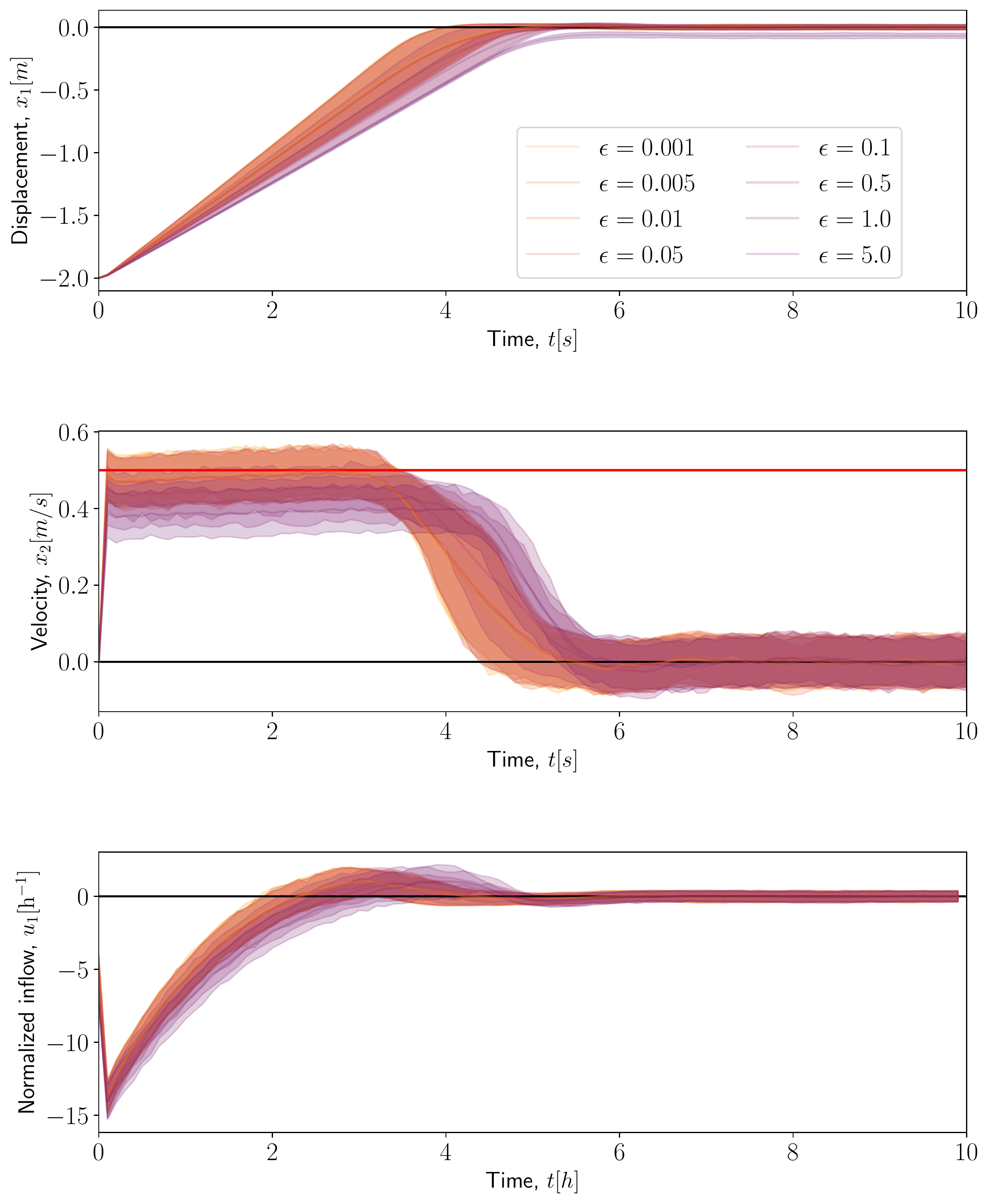}
    \vspace*{-6mm}
    \caption{Simulation results of DRMPC using feedback linearization averaged from 500 realizations with one sample and ball radius ranging from $0.001$ to $5$ on the nonlinear mass spring system. Solid lines are the expected trajectories and  shaded areas represent $15-75\%$ percentile of trajectories.}
      \label{fig:academic_sin_feedback}
  \end{minipage}\hfill
\end{figure}

The first simulation is conducted to illustrate the ability of the proposed DRMPC scheme. We control the nonlinear mass spring system under the additive disturbance $\sin(W)$, where $W \sim \mathcal{N}\left(0, 1\right)$. For both linearization methods, the DRMPC is initialized with a sample, i.e. a disturbance sequence with the length $5$, and run $500$ realizations for each ball radius ranging from $0.001$ to $5.0$. We can read from Fig \ref{fig:academic_sin_suclin} and Fig \ref{fig:academic_sin_feedback} that both methods can control the system to the target state and guarantee the expected system velocity satisfies the prescribed upper bound $0.5\,\text{m/s}$. 

Also, when the ball radius increases, the closed-loop system behaves more conservatively. In the first two rows of Table \ref{tab:results}, the constraint violation's rates of successive linearization between $0.5\,\text{s}$ and $2.0 \, \text{s}$ increase for the both methods when the ball radius increases from $0.001$ to $5.0$.

\subsection{Comparison between two conic formulations applying successive linearization for case study 1}
\label{sec:case_study1_sim2}
The second simulation illustrates the difference between the two conic formulations using successive linearization for the proposed DRMPC scheme. Same as in the first simulation, we control the nonlinear mass spring system under the additive disturbance $\sin(W)$, where $W \sim \mathcal{N}\left(0, 1\right)$. The DRMPC is also initialized with a sample and runs $500$ realizations for each ball radius ranging from $0.001$ to $5.0$. We can read from Fig \ref{fig:two_conic} for the same ball radius, the formulation in \eqref{eq:seq_lin_MPC} behaves less conservative than in \eqref{eq:seq_lin_MPC_new}. The constraint violation's rates of successive linearization using \eqref{eq:seq_lin_MPC_new} between $0.5\,\text{s}$ and $2.0\,\text{s}$ are less than using \eqref{eq:seq_lin_MPC}.

\subsection{Comparison between successive linearization and polynomial chaos SMPC for case study 1}
\label{sec:case_study1_sim3}
\begin{figure}[thpb]
  \begin{minipage}[b]{0.5\textwidth}
  \centering
    \includegraphics[width=\textwidth]{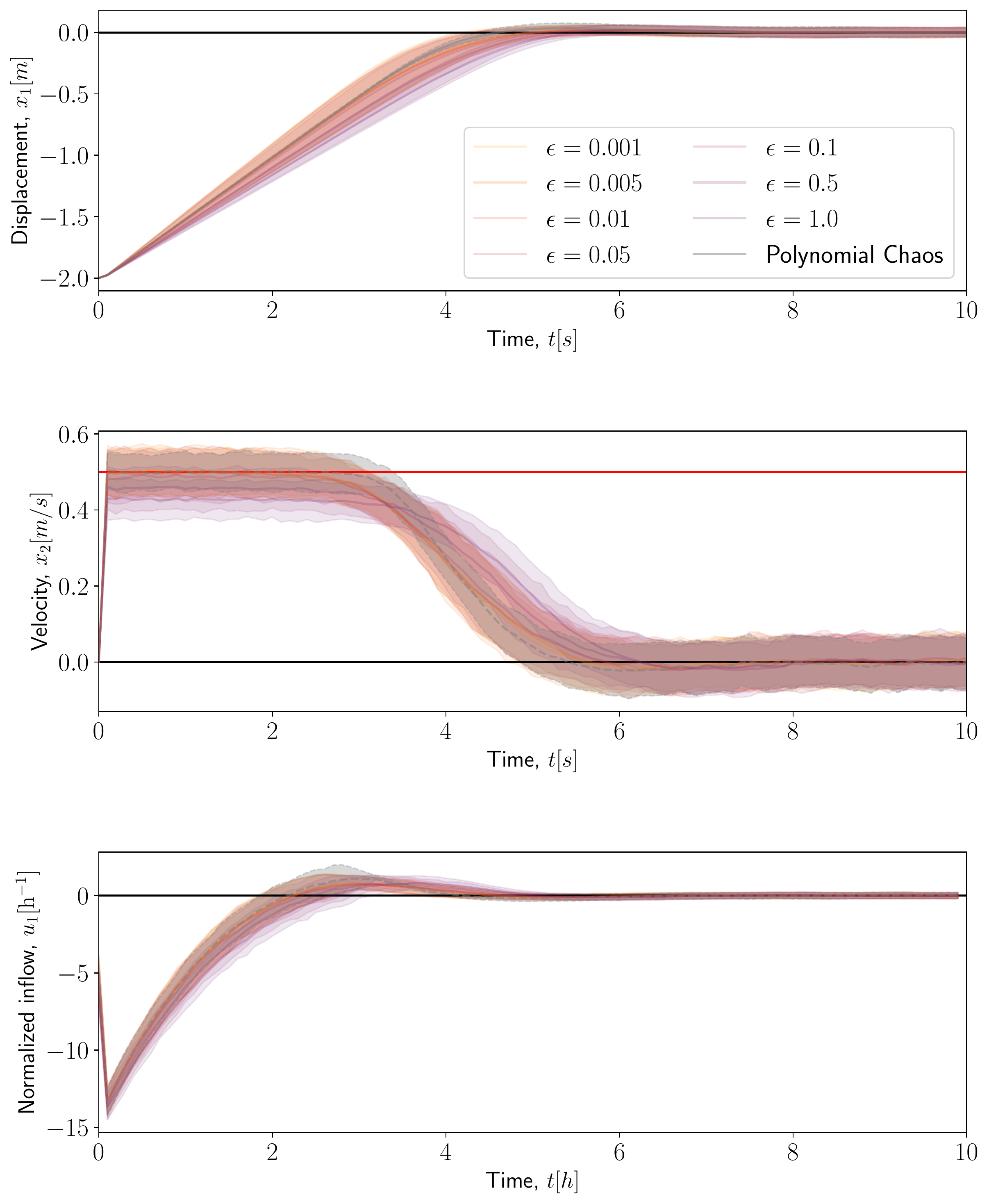}
    \vspace*{-6mm}
    \caption{Simulation results comparing DRMPC using successive linearization and PC-based SMPC in the nominal scenario averaged from 500 realizations with one sample and ball radius ranging from $0.001$ to $5$ on the nonlinear mass spring system. Solid lines are the expected trajectories and  shaded areas represent $15-75\%$ percentile of trajectories.}
    \label{fig:academic_SucL_PolyC}
  \end{minipage}
  \begin{minipage}[b]{0.5\textwidth}
  \centering
    \includegraphics[width=\textwidth]{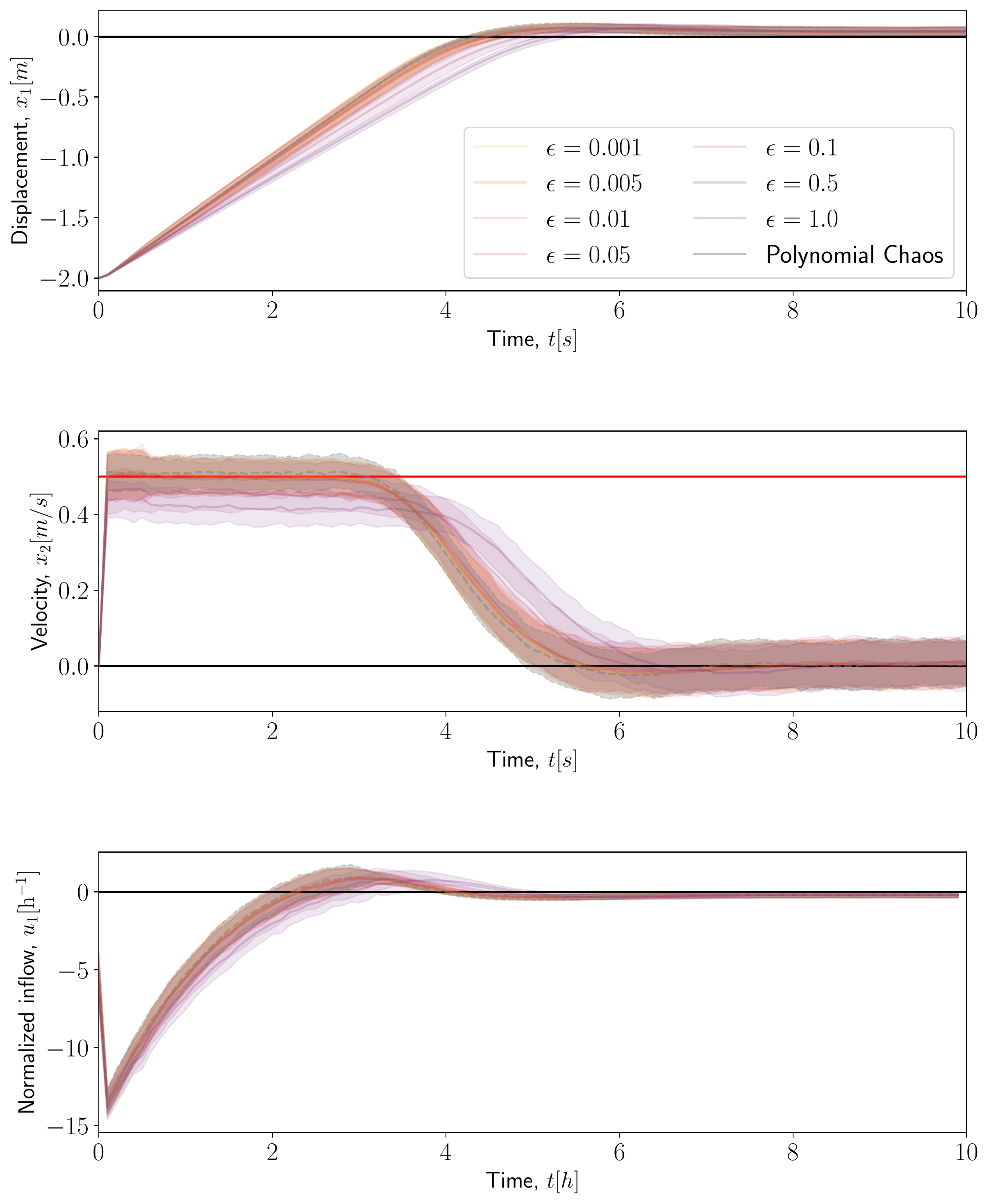}
    \vspace*{-6mm}
    \caption{Simulation results comparing DRMPC using successive linearization and PC-based SMPC under modified distribution  averaged from 500 realizations. DRMPC collects five samples online initially from one sample and ball radius ranges from $0.001$ to $5$ on the nonlinear mass spring system. Solid lines are the expected trajectories and  shaded areas represent $15-75\%$ percentile of trajectories.}
      \label{fig:academic_SucL_Poly_shift}
  \end{minipage}\hfill
\end{figure}

The third simulation compares the proposed DRMPC scheme and a adapted polynomial-chaos-based (PC-based) SMPC in \cite{fagiano2012nonlinear}. To apply polynomial chaos expansions, we assume the system suffers from the uniformly distributed additive disturbance over $[-0.1, 0.1]$ on velocity. For both scheme we run $500$ realizations. The PC-based SMPC collects $25$ additive disturbances offline, whereas the DRMPC collects a disturbance sequence with the length $1$, i.e. consecutive realization of $4$ additive disturbances. We illustrate the simulation results in Fig. \ref{fig:academic_SucL_PolyC}. The constraint violation's rate of PC-based SMPC between $0.5\,\text{s}$ and $2.0\,\text{s}$ is $50\%$.

Now we assume the true distribution deviates from the assume uniform distribution over $[-0.1, 0.1]$. The true additive disturbance is uniformly distributed over $[-0.08, 0.1]$. The constraint violation's rate for PC-based SMPC increases to $55 \%$. However, for DRMPC, we don't assume the exact distribution form. DRMPC is able to guarantee the constraint satisfaction through collecting more samples such that the empirical distribution is close to the true distribution. Under the true additive disturbance,  the constraint violation's rates decrease from $50\%$ to $35\%$, when the ball radius increases to $0.5$. The modified distribution also results in the tracking error when PC-based SMPC is applied. The constraint violation rate is visualized in Fig. \ref{fig:constraint_vio1}.

\subsection{Comparison between successive linearization and polynomial chaos SMPC for case study 2}
\label{sec:case_study2}
\begin{figure}[thpb]
  \begin{minipage}[b]{0.5\textwidth}
  \centering
    \includegraphics[width=\textwidth]{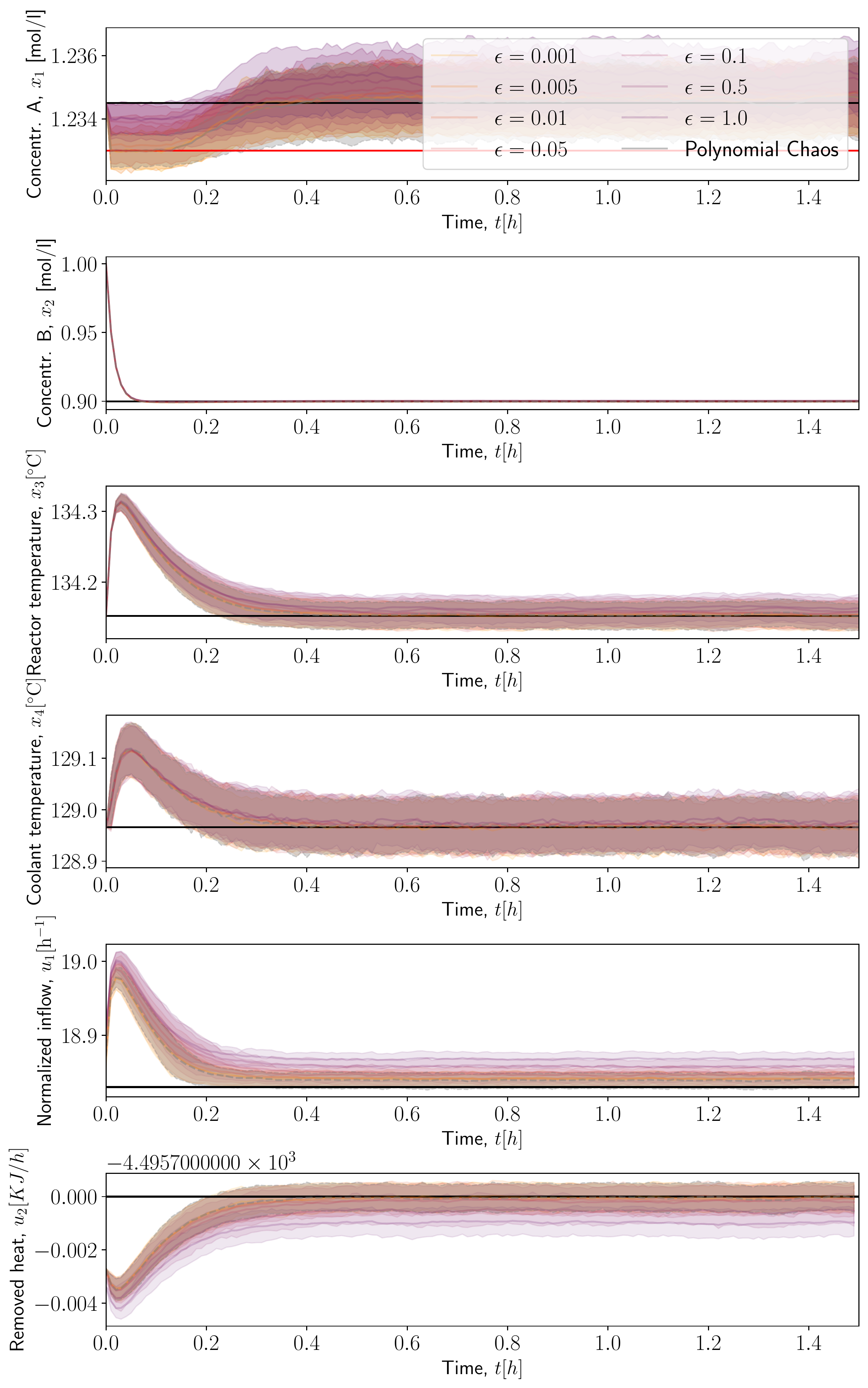}
    \vspace*{-6mm}
    \caption{Simulation results comparing DRMPC using successive linearization and PC-based SMPC in the nominal scenario averaged from 500 realizations with one sample and ball radius ranging from $0.001$ to $5$ on the CSTR system. Solid lines are the expected trajectories and  shaded areas represent $15-75\%$ percentile of trajectories.}
    \label{fig:CSTR_SucL_PolyC}
  \end{minipage}
  \begin{minipage}[b]{0.5\textwidth}
  \centering
    \includegraphics[width=\textwidth]{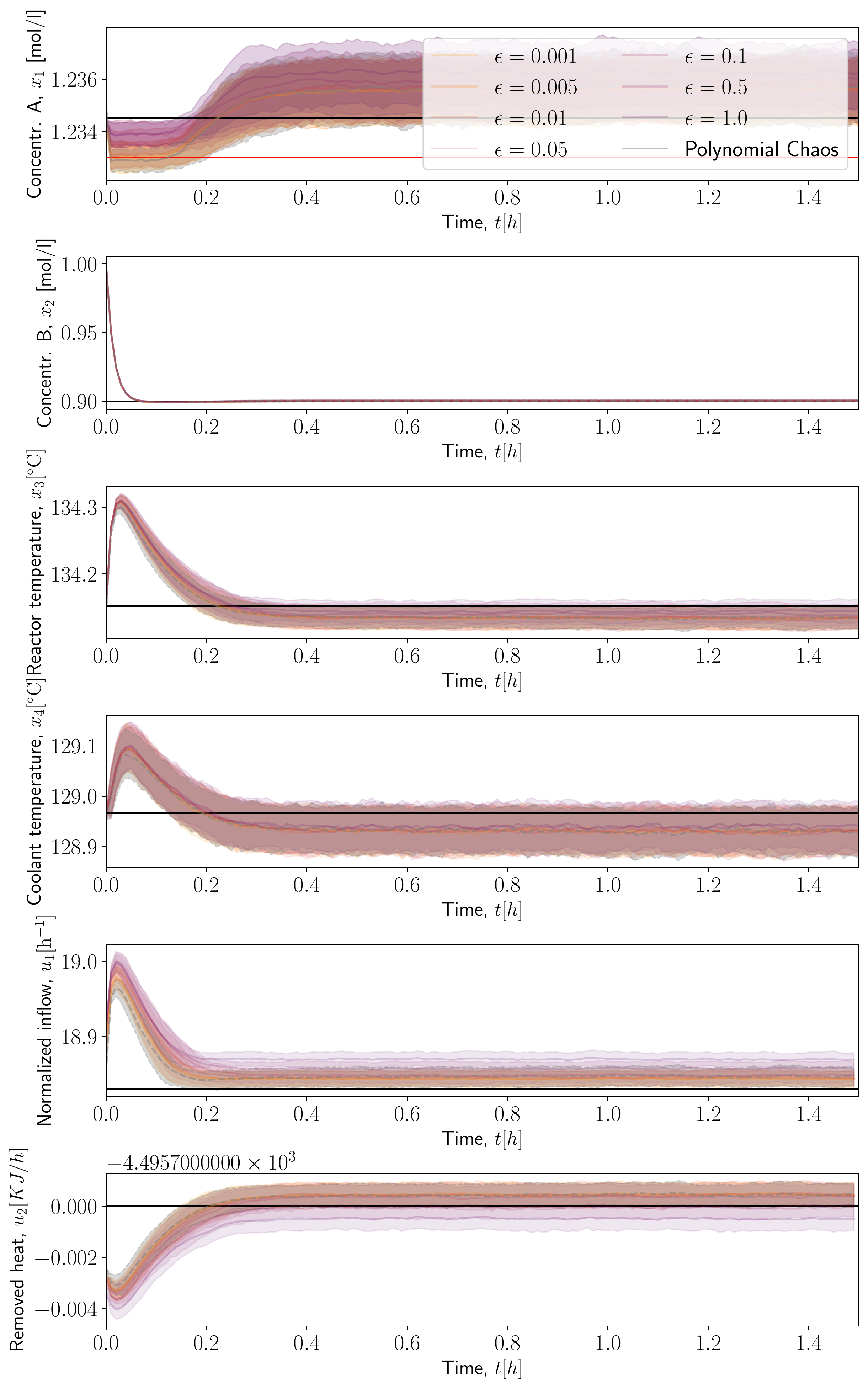}
    \vspace*{-6mm}
    \caption{Simulation results comparing DRMPC using successive linearization and PC-based SMPC under modified distribution  averaged from 500 realizations. DRMPC collects five samples online initially from one sample and ball radius ranges from $0.001$ to $5$ on the CSTR system. Solid lines are the expected trajectories and  shaded areas represent $15-75\%$ percentile of trajectories.}
      \label{fig:CSTR_SucL_Poly_shift}
  \end{minipage}\hfill
\end{figure}

The fourth simulation compares the proposed DRMPC scheme and PC-based SMPC in \cite{fagiano2012nonlinear}. To apply polynomial chaos expansions, we assume the system suffers from the uniformly distributed additive disturbances $w_1$ and $w_2$ over $[-0.001, 0.001]$ and $[-0.1, 0.1]$ on the concentration A $x_1$ and Coolant temperature $x_4$, respectively. For both scheme we run $500$ realizations. The PC-based SMPC collects $40$ additive disturbances offline, whereas the DRMPC collects a disturbance sequence with the length $1$, i.e.  $4$ consecutive realizations for each disturbance $w_1$ and $w_2$. We illustrate the simulation results in Fig. \ref{fig:CSTR_SucL_PolyC}. The constraint violation's rate of PC-based SMPC between $0.01\,\text{s}$ and $0.13\,\text{s}$ is $50\%$.

\begin{figure}[H]
    \begin{minipage}{.32\textwidth}
     \begin{subfigure}{\textwidth}
    \centering
    \includegraphics[width=\textwidth]{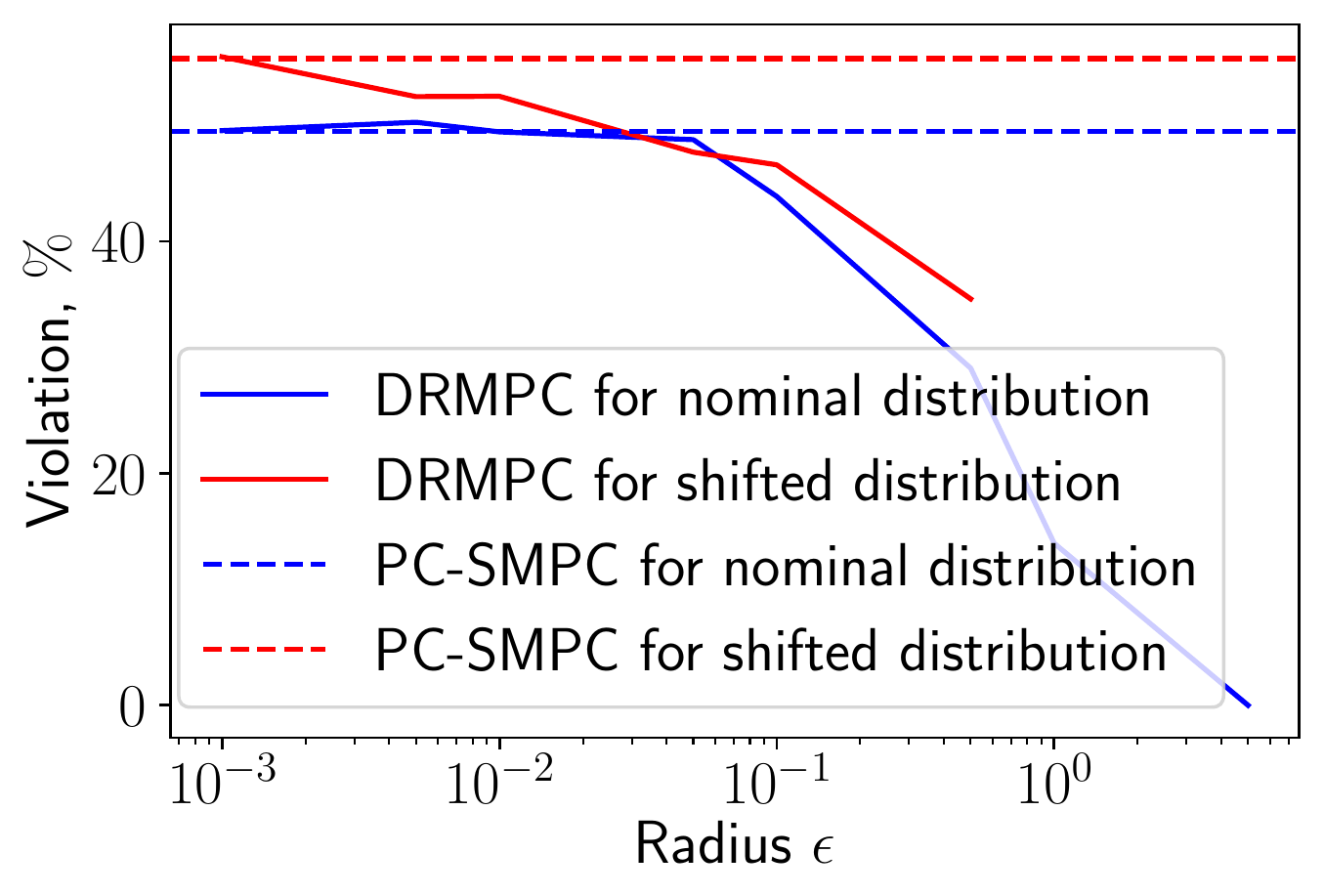}
    \caption{Constraint violation rate in Section \ref{sec:case_study1_sim3}.}
    \label{fig:constraint_vio1}
  \end{subfigure}%
  \end{minipage}
        \hfill  
  \begin{minipage}{.32\textwidth}
  \begin{subfigure}{\textwidth}
    \centering
    \includegraphics[width=\textwidth]{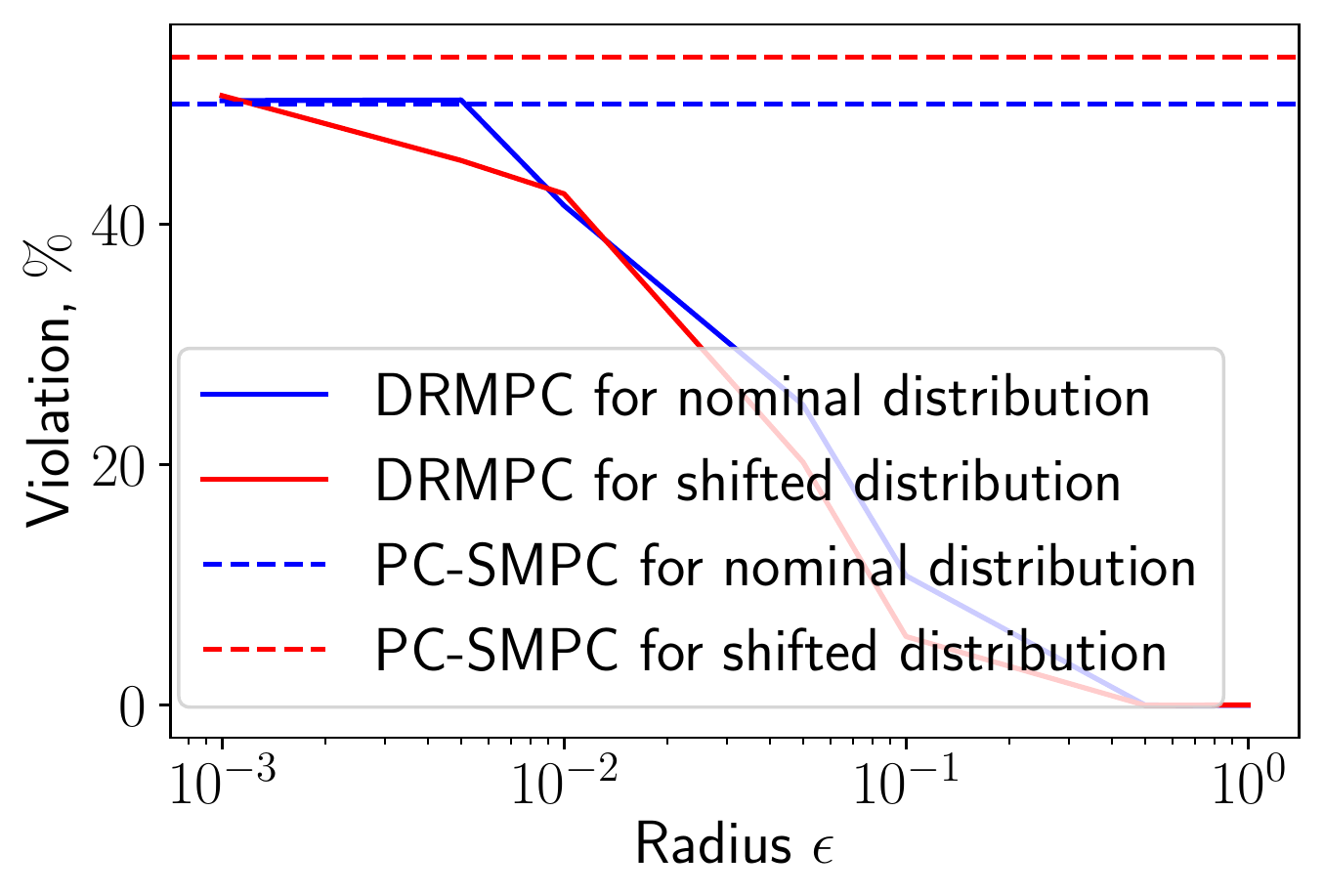}
    \caption{Constraint violation rate in Section \ref{sec:case_study2}.}
      \label{fig:constraint_vio2}
  \end{subfigure}
   \end{minipage}
        \hfill   
   \begin{minipage}{.32\textwidth}
  \begin{subfigure}{\textwidth}\quad
    \centering
    \includegraphics[width=\textwidth]{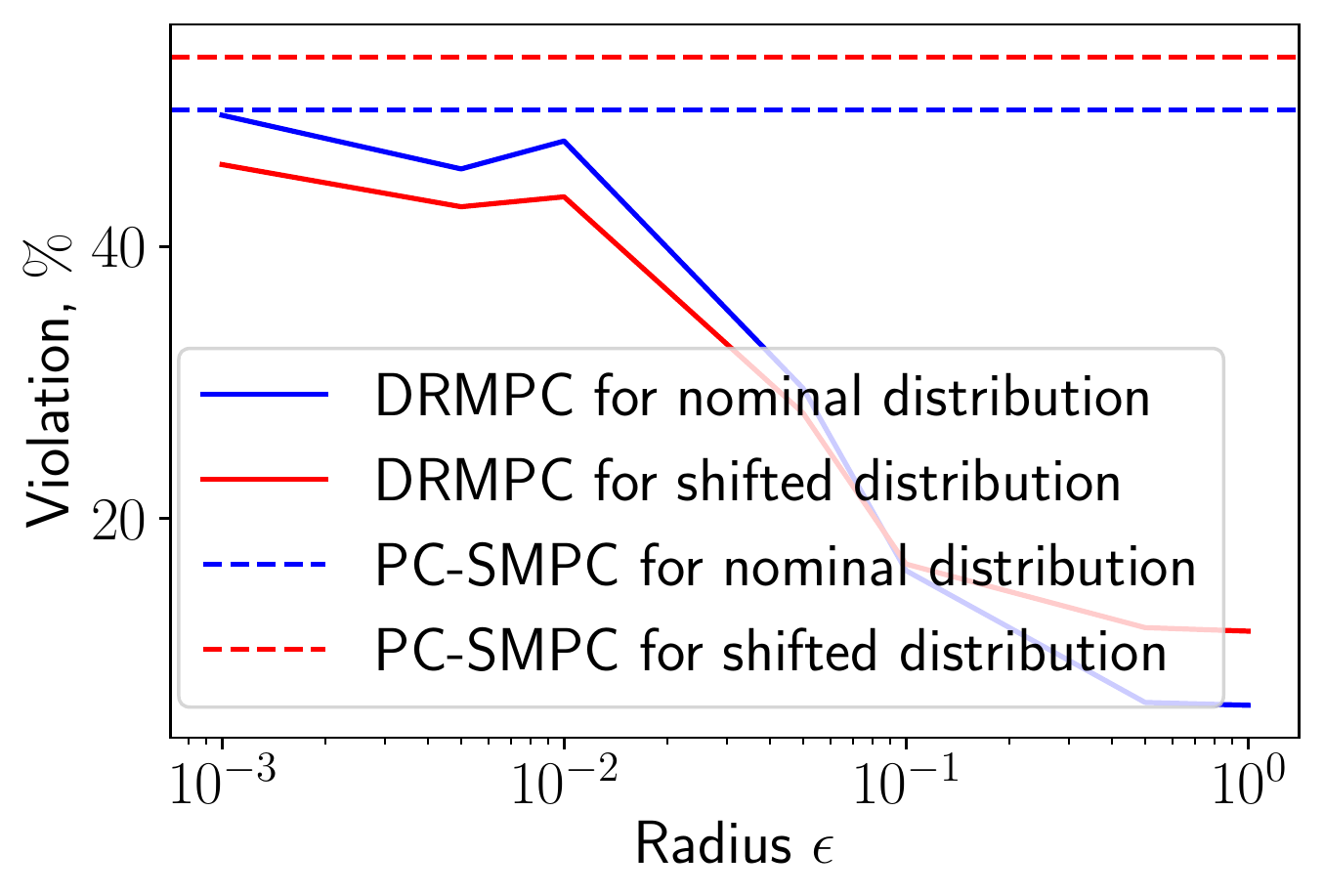}
    \caption{Constraint violation rate in Section \ref{sec:offset_free_simulation}. }
      \label{fig:constraint_vio3}
  \end{subfigure}
  \end{minipage}
  \caption{Relation between the ball radius and constraint violations, averaged from $500$ realizations of trajectories.}
\end{figure}

Now we assume the true distribution deviates from the assumed uniform distribution. The true additive disturbances are uniformly distributed over $[-0.001, 0.0008]$ and $[-0.1, 0.08]$, respectively. The constraint violation's rate for PC-based SMPC increases to $56.33\%$. However, for DRMPC, we do not assume the exact distribution form. DRMPC in Fig. \ref{fig:CSTR_SucL_Poly_shift} is able to guarantee the constraint satisfaction through collecting more samples such that the empirical distribution is close to the true distribution. Under the true additive disturbance, the constraint violation's rates are close to $50\%$ as shown in Table \ref{tab:results}. The modified distribution also results in the tracking error when PC-based SMPC is applied. The constraint violation rate is visualized in Fig. \ref{fig:constraint_vio2}.

\begin{table}
\begin{center}
\begin{tabular}{||p{2.4cm} | c | c | c | c | c  | c | c | c | c ||} 
 \hline
  \backslashbox{Sec.}{Radius} & 1e-3 & 5e-3 & 1e-2 & 5e-2 & 1e-1 & 5e-1 & 1 & 5 & PC-SMPC\\ [0.5ex] 
 \hline\hline
\ref{sec:case_study1_sim1} Suc. Lin.& 50.57\%& 50.95\%& 47.05\%& 47.04\%& 44.00\%& 29.85\%& 14.21\%& 0.05\% & \diagbox[innerwidth=1em, height=\line]{}{}\\ 
 \hline
 \ref{sec:case_study1_sim1} Fee. Lin. & 43.37\%& 40.47 \%&  40.17 \%&   41.03 \%&  40.47 \%&  25.61 \%&  12.05 \%&  0.0\% & \diagbox[innerwidth=1em, height=\line]{}{} \\
 \hline
 \ref{sec:case_study1_sim3} Nom. Dist.& 49.55\%& 50.27\%& 49.44\% & 48.77\%& 43.87\%& 29.07\%& 13.97\%& 0.0\% &49.50\% \\ 
 \hline
 \ref{sec:case_study1_sim3} Shift Dist. & \red{55.91} \%&  \red{52.48}\%&  \red{52.51}\%&  47.69\%&  46.60\%&  35.04\%& \diagbox[innerwidth=1em, height=\line]{}{} & \diagbox[innerwidth=1em, height=\line]{}{} & \red{55.75}\%\\
 \hline
\ref{sec:case_study2} Nom. Dist. & 50.30\%& 50.35\%& 41.57\%& 24.98\%& 10.78\%& 0.02\%& 0.0\%& \diagbox[innerwidth=1em, height=\line]{}{} &49.68\%\\
\hline
  \ref{sec:case_study2} Shift Dist. & 50.71\%&45.32\%& 42.55\%& 20.22\%& 5.72\%& 0.0\%& 0.03\%& \diagbox[innerwidth=1em, height=\line]{}{} &\red{56.33}\%
 \\ [1ex] 
 \hline
 \ref{sec:offset_free_simulation} Nom. Dist. & 49.65\%& 45.70\%& 47.75\%& 29.57\%& 16.17\%& 6.48\%& 6.27\%& \diagbox[innerwidth=1em, height=\line]{}{} & 50.04\%\\
\hline
 \ref{sec:offset_free_simulation} Shift Dist. & 46.02\%&42.92\%& 43.65\%& 27.75\%& 16.65\%& 11.98\%& 11.72\%& \diagbox[innerwidth=1em, height=\line]{}{} & \red{53.94\%}
 \\ [1ex] 
 \hline
 
\end{tabular}
\end{center}

\caption{Constraint violation rates for SMPC with different ball radius and PC-based SMPC for each simulation. Red: The numbers indicate that the expected constraint is not satisfied under the shifted disturbance distribution. The empty blocks of DRMPC with ball radius $1$ or $5$ indicate that the DRMPC becomes infeasible under an unnecessarily large radius. Such a ball radius contains some distributions distant from the true distribution and results in a conservative behavior of the algorithm, similar to the open-loop propagation effect of a large uncertainty set in RMPC \cite{langson2004robust}. }
\label{tab:results}
\end{table}

\subsection{Offset-free tracking}
\label{sec:offset_free_simulation}
As shown in Fig. \ref{fig:CSTR_SucL_Poly_shift}, due to the non-nominal asymmetrical distribution of additive disturbances, there is a contestant tracking error between the expected steady state and tracking point. Similar to Section \ref{sec:case_study2},
we also assume that the true distribution deviates from the nominal uniform distribution. The true additive disturbances are uniformly distributed over $[-0.001, 0.0008]$ and $[-0.1, 0.08]$, respectively.
To address the problem of tracking error, we integrate an offset-free method \cite{morari2012nonlinear} into our framework. More details can be found in \ref{sec:offset-free_append}.
Notice that for DRMPC, we do not assume the exact distribution form. DRMPC in Fig. \ref{fig:CSTR_SucL_Poly_shift} is able to guarantee the constraint satisfaction through collecting more samples such that the empirical distribution is close to the true distribution. Under the true additive disturbance, the feedback linearization the constraint violation's rates are close to $50\%$ as in Table \ref{tab:results}. The modified distribution also results in the tracking error when PC-based SMPC is applied. The constraint violation rate is visualized in Fig. \ref{fig:constraint_vio3}.

\begin{figure}[thpb]
  \begin{minipage}[b]{0.5\textwidth}
  \centering
    \includegraphics[width=\textwidth]{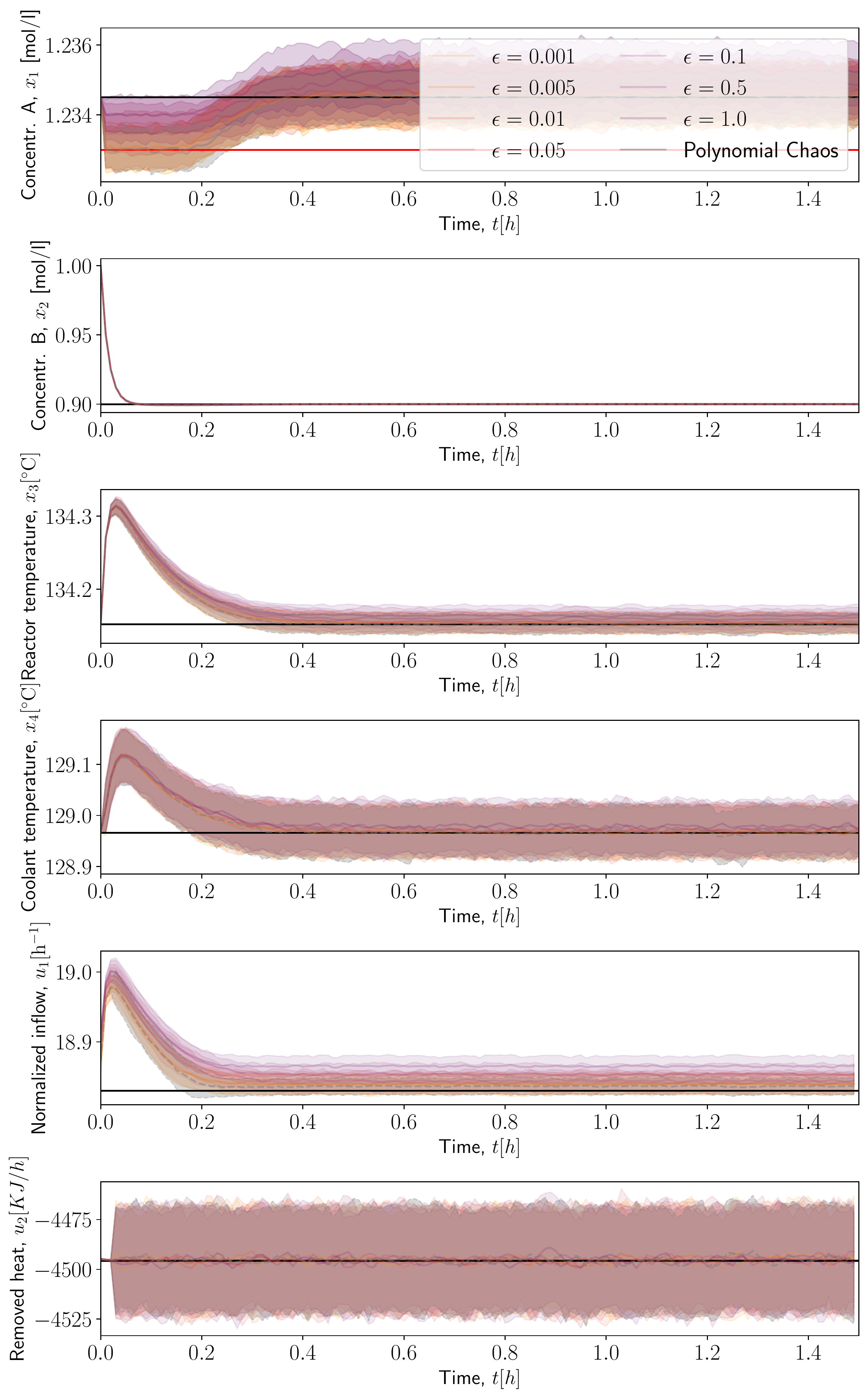}
    \vspace*{-6mm}
    \caption{Simulation results for offset-free DRMPC using successive linearization in the nominal scenario averaged from 500 realizations with one sample and ball radius ranging from $0.001$ to $5$ on the CSTR system. Solid lines are the expected trajectories and  shaded areas represent $15-75\%$ percentile of trajectories.}
    \label{fig:CSTR_SucL_offset_free}
  \end{minipage}
  \begin{minipage}[b]{0.5\textwidth}
  \centering
    \includegraphics[width=\textwidth]{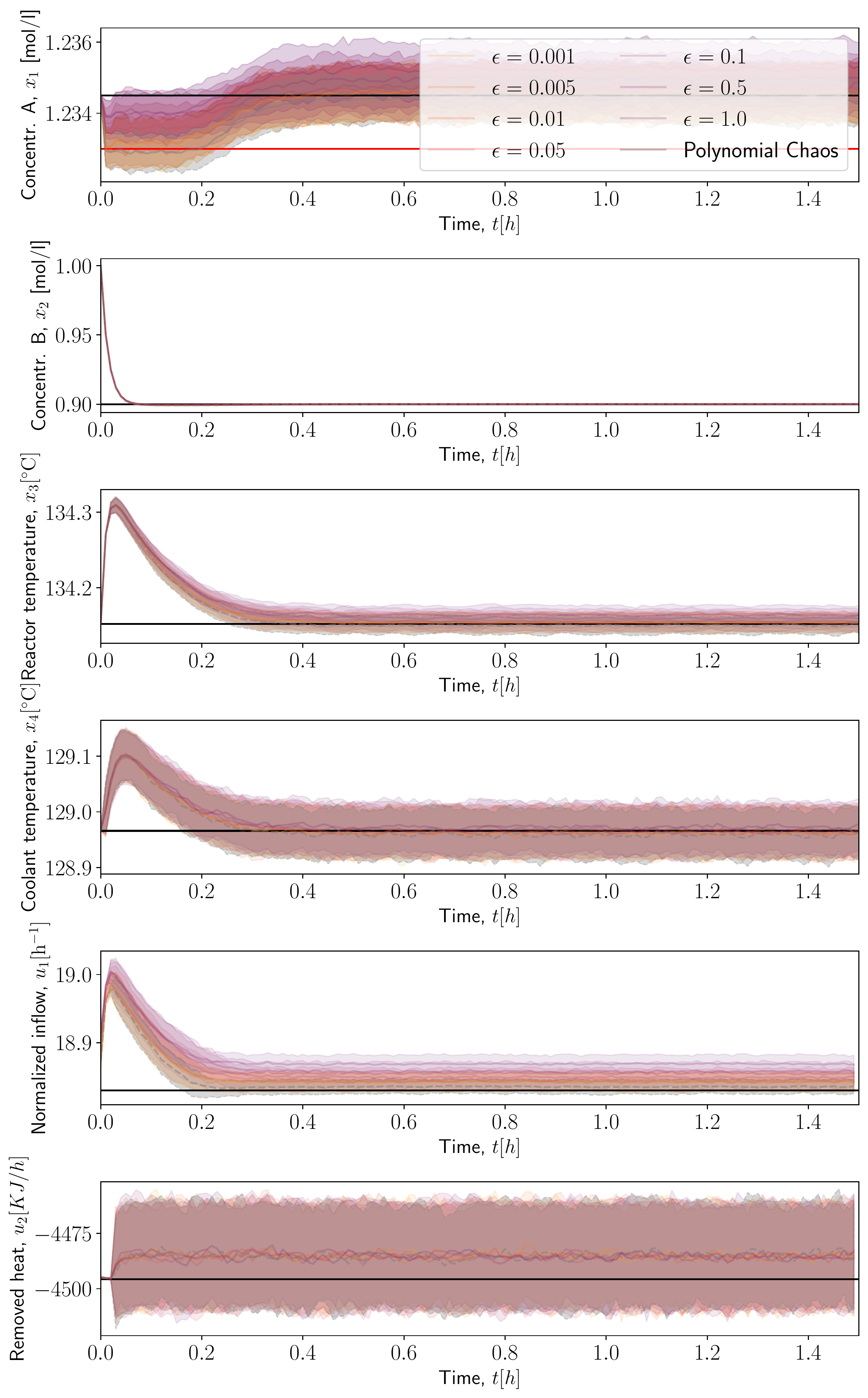}
    \vspace*{-6mm}
    \caption{Simulation results for offset-free DRMPC using successive linearization under modified distribution averaged from 500 realizations with one sample and ball radius ranges from $0.001$ to $5$ on the CSTR system. Solid lines are the expected trajectories and  shaded areas represent $15-75\%$ percentile of trajectories.}
      \label{fig:CSTR_SucL_shifted}
  \end{minipage}\hfill
\end{figure}

\section{Conclusion}
In this paper, we propose a novel data-driven DRMPC scheme for nonlinear systems with additive disturbance using  the Wasserstein ambiguity. This scheme is proposed to guarantee that the nominal constraints are satisfied in expectation, in a distributionally robust sense. Successive linearization and feedback linearization are applied to approximate the prototype DRMPC scheme. The approximated control scheme can be efficiently solved through an equivalent conic program. Numerical case studies on a nonlinear mass spring system and CSTR system are conducted to illustrate the effectiveness of the scheme over the classic polynomial chaos-based SMPC. Future work will focus on developing a scheme with chance constraints while preserving computational tractability.

\appendix
\subsection{Offset-free tracking}
\label{sec:offset-free_append}
In this subsection, we incorporate an offset-free method to address the problem of tracking error. We consider integrating the method proposed in \cite{morari2012nonlinear} into our DRMPC framework. The idea behind offset-free reference tracking is augmenting the nominal model with a disturbance state $d$, such that the side-effect of model mismatch or exogenous disturbances can be captured and reduced by this augmented disturbance state. For such an augmented system, 
\renewcommand{\theequation}{A.\arabic{equation}}
\begin{equation}
\label{eq:aug_sys}
\begin{aligned}
&x(k+1)=f_{\text {aug }}(x(k), d(k), u(k)) \\
&d(k+1)=d(k) \\
&y(k)=g_{\text {aug }}(x(k), d(k)),
\end{aligned}
\end{equation}
system state $\hat{x}(k+1)$ and disturbance $\hat{d}(k)$ are estimated via an observer at each sampling time, so that a new target point ($\bar{x}, \bar{u}$) (decision variables) as an equilibrium point for the augmented system can be updated according to the estimation, i.e. $\bar{x}=f_{\text {aug }}\left(\bar{x}, \hat{d}(k), \bar{u}\right)$. In this work, the output $y$ indicates the elements of state, of which the tracking error between $y$ and the reference $r$ should be reduced.

Here we mainly focus on the design of the CSTR system \eqref{eq:CSTR_ODE}. However, this method can be generally applied to other nonlinear systems. First, we define the augmented system as 
\begin{equation}
\label{eq:aug_sys}
\begin{aligned}
&x(k+1)=f_{\text{CSTR}}(x(k), u(k))+ \begin{bmatrix}
d_1(k)\\d_2(k)\\0\\0
\end{bmatrix} \\
&d(k+1)=d(k):= \begin{bmatrix}
d_1(k)\\d_2(k)
\end{bmatrix}\\
&y(k)=\begin{bmatrix}
1 & 0 & 0 & 0\\
0 & 0 & 1 & 0\\
\end{bmatrix}x(k),
\end{aligned}
\end{equation}
where $f_{\text{CSTR}}$ is the discretized nominal CSTR system \eqref{eq:CSTR_ODE}. Then, we define the observer applied in this case study as 
\begin{equation}
\begin{aligned}
\hat{x}(k+1)=& f_{\text {CSTR }}(\hat{x}(k),u(k)) + \begin{bmatrix}
\hat{d}_1(k)\\\hat{d}_2(k)\\0\\0
\end{bmatrix} +L_x\left(y_{\text{real}} - \begin{bmatrix}
1 & 0 & 0 & 0\\
0 & 0 & 1 & 0\\
\end{bmatrix}\hat{x}(k)\right) \\
\hat{d}(k+1)=& \hat{d}(k)+L_d\left(y_{\text{real}} - \begin{bmatrix}
1 & 0 & 0 & 0\\
0 & 0 & 1 & 0\\
\end{bmatrix}\hat{x}(k)\right),
\end{aligned}
\end{equation}
where $L_x = \begin{bmatrix}
0.98475743, & -0.03074359 \\
0.18526631, & 0.75600808 \\
1.16506026, & 1.67394808 \\
0.43648359, & 1.03137772 \\
\end{bmatrix}$ and $L_d = \begin{bmatrix}
0.59509589, & 0.01123627 \\
-0.02060668, & 0.39146257
\end{bmatrix}$ are designed based on the discretized nominal model $f_{\text{CSTR}}$ around $x_r$ and $u_r$, using the steady-state Kalman filter algorithm with $Q_w = 10\mathbf{I}$ and $R_v = \mathbf{I}$, see \cite[Eq. 16]{pannocchia2003disturbance}. After acquiring the observer, we extend the optimization problem \eqref{eq:approx_prob} with equality conditions adaptively identifying tracking points for the disturbed system. Together with the extra conditions, we apply the estimated disturbance and state in the following optimal control problem for the purpose of offset-free tracking:
\begin{equation}
\label{eq:offset-tracking-DRMPC}
\begin{array}{cl}
\displaystyle\min_{\mathbf{z}, \mathbf{v}, \bar{x}, \bar{u}} & \mathbb{E}_{\mathbb{P}} \{\sum_{i=0}^{N-1} (\|z_{i\mid k} + e_{i\mid k} - \bar{x} \|_{Q}^{2}  + \|v_{i\mid k} + Ke_{i\mid k} - \bar{u}\|_{R}^{2}) +  \| z_{N\mid k} + e_{N\mid k} - \bar{x}\|_{Q_f}^{2})\} \\
\text { s.t. } 
& z_{0 \mid k}=\hat{x}_{k}\\
& e_{0 \mid k} \stackrel{a . s .}{=} 0. \\
& \bar{x}=A_k \bar{x}+ B_k \bar{u} + \delta_k + \begin{bmatrix}
\hat{d}_1(k)\\\hat{d}_2(k)\\0\\0
\end{bmatrix}\\
&r =  \begin{bmatrix}
1 & 0 & 0 & 0\\
0 & 0 & 1 & 0\\
\end{bmatrix} \bar{x}\\
& z_{i+1 \mid k}=A_k z_{i \mid k}+ B_k v_{i \mid k} + \delta_k  +  \begin{bmatrix}
\hat{d}_1(k)\\\hat{d}_2(k)\\0\\0
\end{bmatrix}\\
& e_{i+1 \mid k}=A^{cl}_{k}  e_{i \mid k}+D W_{i+k} \\ 
& \displaystyle\sup_{\mathbb{P}_k \in \mathcal{P}_k} \mathbb{E}_{\mathbb{P}_k}\left\{[F]_{j}(z_{i\mid k} + e_{i\mid k} + \delta_k)\right\}  \leq [f]_{j}, \quad k \in \mathbb{N}_{\ge 0}, j \in \mathbb{N}_1^{n_F}, i \in \mathbb{N}_1^N \\
&\displaystyle \max_{e_{i \mid k}} G (K e_{i \mid k}+v_{i \mid k})  \leq g \quad  i \in \mathbb{N}_0^{N-1},\\
\end{array}
\end{equation}
where $r = [1.2345, 134.15]^{\top}$. The optimization problem $\eqref{eq:offset-tracking-DRMPC}$ can be reformulated exactly into a conic program 
\begin{equation}
\label{eq:offset_lin_MPC}
\begin{array}{cl}
\displaystyle\min_{
\substack{\boldsymbol{z}, \boldsymbol{v}, \bar{x}, \bar{v}, \boldsymbol{x}_{\text{aux}}, \boldsymbol{u}_{\text{aux}}\\
    \gamma \ge 0, \lambda \ge 0, s, \xi_{dual1}\ge 0}}
      & \sum_{i=1}^{N-1} x_{\text{aux}, i}^{\top} Q x_{\text{aux}, i}+u_{\text{aux}, i}^{\top} R u_{\text{aux}, i}+x_{\text{aux}, N}^{\top} Q x_{\text{aux}, N} \\
\text { s.t. } 
& z_{0\mid k}=\hat{x}_{k}, \\
& z_{i+1 \mid k}=A_k z_{i \mid k}+ B_k v_{i \mid k} + \delta_k  +  \begin{bmatrix}
\hat{d}_1(k)\\\hat{d}_2(k)\\0\\0
\end{bmatrix}\\
& x_{\text{aux}, i}  = z_{i \mid k} - \bar{x}\\
& u_{\text{aux}, i}  = v_{i \mid k} - \bar{u}\\
& \bar{x}=A_k \bar{x}+ B_k \bar{u} + \delta_k + \begin{bmatrix}
\hat{d}_1(k)\\\hat{d}_2(k)\\0\\0
\end{bmatrix}\\
&r =  \begin{bmatrix}
1 & 0 & 0 & 0\\
0 & 0 & 1 & 0\\
\end{bmatrix} \bar{x}\\
\begin{aligned}
i & \in \mathbb{N}_0^{N-1}\\  j & \in \mathbb{N}_1^{n_G}
\end{aligned}
&
\left\{\begin{aligned}
&h^{\top} \xi_{\text {dual } 1, i j} \leq I_{j}^{\top}\left(g-G v_{i}\right)\\
&H^{\top} \xi_{dual1,ij} = L_{ij}\\
& \xi_{dual1,ij} \ge 0
\end{aligned}\right.\\
\begin{aligned}
i & \in \mathbb{N}_1^{N-1}\\  j & \in \mathbb{N}_1^{n_F}
\end{aligned}
&
\left\{\begin{aligned}
& \lambda_{ij} \varepsilon + \frac{1}{N_k} \sum_{l=1}^{N_k}s_{ijl}  \le 0\\
& I_j^{\top}(F (z_{i} +[D_x]_{i \times n_x + 1: (i+1) \times n_x }\hat{\xi}^{(l)}) - f )+ \gamma_{ijl}^{\top}(h_{\xi}-H_{\xi} \hat{\xi}^{(l)})  \le s_{ijl}\\
& \|H_{\xi}^{\top}\gamma_{ijl} -  I_j^{\top}(F[D_x]_{i \times n_x + 1: (i+1) \times n_x}) \|_{\infty} \le \lambda_{ij}\\
& \gamma_{ijl}  \ge 0\\
\end{aligned}\right.\\
\begin{aligned}
i &= N\\  j & \in \mathbb{N}_1^{n_{F_N}}
\end{aligned}
&
\left\{
\begin{aligned}
&\lambda_{ij} \varepsilon + \frac{1}{N_k} \sum_{l=1}^{N_k}s_{ijl}  \le 0\\
& I_j^{\top}(F_N (z_{i}+[D_x]_{i \times n_x + 1: (i+1) \times n_x }\hat{\xi}^{(l)}) - f_N)+ \gamma_{ijl}^{\top} (h_{\xi}-H_{\xi} \hat{\xi}^{(l)} ) \le s_{ijl}\\
& \|H_{\xi}^{\top}\gamma_{ijl} -  I_j^{\top}(F_N[D_x]_{i \times n_x + 1: (i+1) \times n_x }) \|_{\infty}  \le \lambda_{ij}\\
& \gamma_{ijl}  \ge 0
\end{aligned}\right. \\
& l \in \mathbb{N}_1^{N_k}
\end{array}
\end{equation}
by leveraging same techniques in Theorem \ref{th:exact_reform}. In conclusion, compared to the original DRMPC, the offset-free method requires the estimation of state and disturbance at each sampling time, and such an estimation is used in the modified optimization problem to shrink the tracking error adaptively.

\bibliography{example}

\end{document}